\documentclass[longauth]{aa}

\usepackage{graphicx}
\usepackage{txfonts}
\usepackage{mathabx}
\usepackage{placeins}
\bibpunct{(}{)}{;}{a}{}{,} 
\def \MJ{M$_{\mathrm{J}}$}
\def \ME{M$_{\Earth}$}
\def \RJ{R$_{\mathrm{J}}$}
\def \RE{R$_{\Earth}$}
\def \RS{R$_{\odot}$}
\def \msol{M$\mathrm{_\odot}$}

\def \kms{km\,s$^{-1}$}
\def \ms{m\,s$^{-1}$}
\def \1s{$1\,\sigma$}
\def \t0{T$_0$}

\def \gcc{g\,cm$^{-3}$}
\usepackage{hyperref}

\begin{document}

    \title{TOI-1199\:b and TOI-1273\:b: Two new transiting hot Saturns
    detected and characterized with SOPHIE and TESS}
    \titlerunning{Detection and characterization of TOI-1199~b and TOI-1273~b}
    
\author{
J. Serrano Bell \inst{1,2} $^{\href{https://orcid.org/0000-0002-8397-557X}{\includegraphics[scale=0.5]{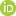}}}$ \and
R. F. D\'iaz\inst{1} $^{\href{https://orcid.org/0000-0001-9289-5160}{\includegraphics[scale=0.5]{img/orcid.jpg}}}$  \and
G. H\'ebrard\inst{2,3}  \and
E. Martioli\inst{4,2} $^{\href{https://orcid.org/0000-0002-5084-168X}{\includegraphics[scale=0.5]{img/orcid.jpg}}}$ \and
N. Heidari\inst{2} \and
S. Sousa\inst{5} $^{\href{https://orcid.org/0000-0001-9047-2965}{\includegraphics[scale=0.5]{img/orcid.jpg}}}$ \and
I. Boisse\inst{6} \and
J. M. Almenara\inst{7} \and 
J. Alonso-Santiago\inst{8} $^{\href{https://orcid.org/0000-0001-9707-3107}{\includegraphics[scale=0.5]{img/orcid.jpg}}}$ \and 
S. C. C. Barros\inst{5} $^{\href{https://orcid.org/0000-0003-2434-3625}{\includegraphics[scale=0.5]{img/orcid.jpg}}}$ \and
P. Benni\inst{9} $^{\href{https://orcid.org/0000-0001-6981-8722}{\includegraphics[scale=0.5]{img/orcid.jpg}}}$ \and
A. Bieryla\inst{10} $^{\href{https://orcid.org/0000-0001-6637-5401}{\includegraphics[scale=0.5]{img/orcid.jpg}}}$  \and 
X. Bonfils\inst{7} \and
D. A. Caldwell\inst{11} $^{\href{https://orcid.org/0000-0003-1963-9616}{\includegraphics[scale=0.5]{img/orcid.jpg}}}$ \and
D. R. Ciardi\inst{12} \and
K. A.\ Collins\inst{13} $^{\href{https://orcid.org/0000-0001-6588-9574}{\includegraphics[scale=0.5]{img/orcid.jpg}}}$ \and
P. Cortés-Zuleta\inst{6} \and
S. Dalal\inst{14,2} \and
J. P. de Le\'on\inst{15} $^{\href{https://orcid.org/0000-0002-6424-3410}{\includegraphics[scale=0.5]{img/orcid.jpg}}}$ \and 
M. Deleuil\inst{6} \and
X. Delfosse\inst{7} \and
O. D. S. Demangeon\inst{5} $^{\href{https://orcid.org/0000-0001-7918-0355}{\includegraphics[scale=0.5]{img/orcid.jpg}}}$ \and
E. Esparza-Borges\inst{16,17} $^{\href{https://orcid.org/0000-0002-2341-3233}{\includegraphics[scale=0.5]{img/orcid.jpg}}}$ \and
T. Forveille\inst{7} $^{\href{https://orcid.org/0000-0003-0536-4607}{\includegraphics[scale=0.5]{img/orcid.jpg}}}$ \and
A. Frasca\inst{8} $^{\href{https://orcid.org/0000-0002-0474-0896}{\includegraphics[scale=0.5]{img/orcid.jpg}}}$ \and 
A. Fukui\inst{18} $^{\href{https://orcid.org/0000-0002-4909-5763}{\includegraphics[scale=0.5]{img/orcid.jpg}}}$ \and 
J. Gregorio\inst{19} $^{\href{https://orcid.org/0000-0002-0145-5248}{\includegraphics[scale=0.5]{img/orcid.jpg}}}$ \and
N. M.\ Guerrero\inst{20,21} \and
S. B. Howell\inst{22} $^{\href{https://orcid.org/0000-0002-2532-2853}{\includegraphics[scale=0.5]{img/orcid.jpg}}}$ \and 
S. Hoyer\inst{6} $^{\href{https://orcid.org/0000-0003-3477-2466}{\includegraphics[scale=0.5]{img/orcid.jpg}}}$ \and
K. Ikuta\inst{15} $^{\href{https://orcid.org/0000-0002-5978-057X}{\includegraphics[scale=0.5]{img/orcid.jpg}}}$ \and
J. M. Jenkins\inst{22} $^{\href{https://orcid.org/0000-0002-4715-9460}{\includegraphics[scale=0.5]{img/orcid.jpg}}}$ \and 
F. Kiefer\inst{2,23} \and
D. W.\ Latham\inst{10} \and 
G. Marino\inst{24,25} $^{\href{https://orcid.org/0000-0001-8134-0389}{\includegraphics[scale=0.5]{img/orcid.jpg}}}$ \and
E. J. Michaels\inst{26} \and
C. Moutou\inst{27} \and
F. Murgas\inst{16,17} $^{\href{https://orcid.org/0000-0001-9087-1245}{\includegraphics[scale=0.5]{img/orcid.jpg}}}$ \and
N. Narita\inst{18,28,16} $^{\href{https://orcid.org/0000-0001-8511-2981}{\includegraphics[scale=0.5]{img/orcid.jpg}}}$ \and
E. Palle\inst{16,17} $^{\href{https://orcid.org/0000-0003-0987-1593}{\includegraphics[scale=0.5]{img/orcid.jpg}}}$ \and
H. Parviainen\inst{17,16} $^{\href{https://orcid.org/0000-0001-5519-1391}{\includegraphics[scale=0.5]{img/orcid.jpg}}}$ \and
N. C. Santos\inst{5,29} \and
K. G.\ Stassun\inst{30} $^{\href{https://orcid.org/0000-0002-3481-9052}{\includegraphics[scale=0.5]{img/orcid.jpg}}}$ \and
J. N.\ Winn\inst{31} $^{\href{https://orcid.org/0000-0002-4265-047X}{\includegraphics[scale=0.5]{img/orcid.jpg}}}$
}

\institute{
\inst{1} International Center for Advanced Studies (ICAS) and ICIFI (CONICET), ECyT-UNSAM, Campus Miguelete, 25 de Mayo y Francia, (1650) Buenos Aires, Argentina. \\
\inst{2} Institut d'astrophysique de Paris, CNRS, UMR 7095, Sorbonne Universit\'{e}, 98 bis bd Arago, 75014 Paris, France. \\
\inst{3} Observatoire de Haute-Provence, St Michel l'Observatoire, France. \\
\inst{4} Laborat\'orio Nacional de Astrofisica, Rua Estados Unidos 154, 37504-364, Itajubá - MG, Brazil \\
\inst{5} Instituto de Astrof{\'\i}sica e Ci\^encias do Espa\c{c}o, Universidade do Porto, CAUP, Rua das Estrelas, 4150-762 Porto, Portugal \\
\inst{6} Laboratoire d'astrophysique de Marseille, Univ. de Provence, UMR6110 CNRS, 38 r. F. Joliot Curie, 13388 Marseille cedex 13, France \\
\inst{7} Univ. Grenoble Alpes, CNRS, IPAG, 414 rue de la Piscine, 38400 St-Martin d’Hères, France \\
\inst{8} INAF - Osservatorio Astrofisico di Catania, Via S. Sofia 78, 95123 Catania, Italy \\
\inst{9} Acton Sky Portal private observatory, Acton, MA, USA \\
\inst{10} Center for Astrophysics \textbar \ Harvard \& Smithsonian, 60 Garden Street, Cambridge, MA 02138, USA \\
\inst{11} SETI Institute/NASA Ames Research Center, Moffett Field, CA 94035, USA \\
\inst{12} IPAC-NASA Exoplanet Science Institute, 770 S. Wilson Avenue, Pasadena, CA 91106, USA \\
\inst{13} Center for Astrophysics \textbar \ Harvard \& Smithsonian, 60 Garden Street, Cambridge, MA 02138, USA \\
\inst{14} Astrophysics Group, University of Exeter, Exeter EX4 2QL, UK \\
\inst{15} Department of Multi-Disciplinary Sciences, Graduate School of Arts and Sciences, The University of Tokyo, 3-8-1 Komaba, Meguro, Tokyo 153-8902, Japan \\
\inst{16} Instituto de Astrof\'isica de Canarias (IAC), calle V\'ia L\'actea s/n, 38205 La Laguna, Tenerife, Spain \\
\inst{17} Departamento de Astrof\'isica, Universidad de La Laguna (ULL), 38206 La Laguna, Tenerife, Spain \\
\inst{18} Komaba Institute for Science, The University of Tokyo, 3-8-1 Komaba, Meguro, Tokyo 153-8902, Japan \\
\inst{19} Crow Observatory, Portalegre, Portugal \\
\inst{20} Bryant Space Science Center, Department of Astronomy, University of Florida, Gainesville, FL 32611, USA \\
\inst{21} Department of Physics and Kavli Institute for Astrophysics and Space Research, Massachusetts Institute of Technology, Cambridge, MA 02139, USA \\
\inst{22} NASA Ames Research Center, Moffett Field, CA 94035, USA \\
\inst{23} LESIA, Observatoire de Paris, Universit\'e PSL, CNRS, Sorbonne Universit\'e, Universit\'e Paris Cit\'e, 5 place Jules Janssen, 92195 Meudon, France \\
\inst{24} Wild Boar Remote Observatory, San Casciano in val di Pesa, Firenze, Italy \\
\inst{25} Gruppo Astrofili Catanesi, Catania, Italy \\
\inst{26} Waffelow Creek Observatory, 10780 FM 1878, Nacogdoches, TX 75961, USA \\
\inst{27} Universit\'e de Toulouse, CNRS, IRAP, 14 avenue Belin, 31400 Toulouse, France \\
\inst{28} Astrobiology Center, 2-21-1 Osawa, Mitaka, Tokyo 181-8588, Japan \\
\inst{29} Departamento de F\'isica e Astronomia, Faculdade de Ci\^encias, Universidade do Porto, Rua do Campo Alegre, 4169-007 Porto, Portugal \\ 
\inst{30} Department of Physics and Astronomy, Vanderbilt University, Nashville, TN 37235, USA \\ 
\inst{31} Department of Astrophysical Sciences, Princeton University, Princeton, NJ 08544, USA \\ }

\abstract{We report the characterization of two planet candidates detected by the Transiting Exoplanet Survey Satellite (TESS), TOI-1199\:b and TOI-1273\:b, with periods of 3.7 and 4.6\,days, respectively. Follow-up observations for both targets, which include several ground-based light curves, confirmed the transit events. High-precision radial velocities from the SOPHIE spectrograph revealed signals at the expected frequencies and phases of the transiting candidates and allowed mass determinations with a precision of 8.4\% and 6.7\% for TOI-1199\:b and TOI-1273\:b, respectively. 
The planetary and orbital parameters were derived from a joint analysis of the radial velocities and photometric data. We find that the planets have masses of 0.239$\,\pm\,$0.020\,\MJ \ and 0.222$\,\pm\,$0.015\,\MJ \ and radii of 0.938$\,\pm\,$0.025\,\RJ \ and 0.99$\,\pm\,$0.22\,\RJ,\ respectively. The grazing transit of TOI-1273\:b translates to a larger uncertainty in its radius, and hence also in its bulk density, compared to TOI-1199\:b. The inferred bulk densities of 0.358$\,\pm\,$0.041\,\gcc \ and 0.28$\,\pm\,$0.11\,\gcc \ are among the lowest known for exoplanets in this mass range, which, considering the brightness of the host stars ($V$$\approx$11\,mag), render them particularly amenable to atmospheric characterization via the transit spectroscopy technique. The better constraints on the parameters of TOI-1199\:b provide a transmission spectroscopy metric of 134\,$\pm$\,17, making it the better suited of the two planets for atmospheric studies.
}
\maketitle
\section{Introduction}
The search for exoplanets has benefited greatly from space-based transit surveys: many thousands of planet-hosting candidates have been found by looking for small periodic dips in the brightness of stars. Nevertheless, additional observations are needed to discard false positives and characterize the systems. The Transiting Exoplanet Survey Satellite (TESS) was launched in September 2018 with the objective of detecting exoplanets around bright nearby stars by monitoring the flux variations of hundreds of thousands of stars in the solar neighborhood \citep{2015JATIS...1a4003R}, covering $\sim$85\% of the sky. One key aspect of the mission is that its targets are relatively bright, making them amenable to follow up with precise radial velocity (RV) observations from the ground and potential atmospheric studies with the\textit{ James Webb }Space Telescope and other facilities. Since TESS observes each sector for $\sim$27\,days, the planets found so far mostly have short periods (<\,30\,days) but a wide range of masses and radii. Therefore, RV surveys are important not only for determining masses but also for providing a way to detect and study long-period systems.

At the time of writing, more than 400 planet candidates from TESS have been confirmed and there are thousands still awaiting their nature to be determined\footnote{\url{https://exoplanetarchive.ipac.caltech.edu/}}. A community effort is being made through the TESS Follow-up Observing Program  \citep[TFOP;][]{2018AAS...23143908C} to efficiently carry out detailed analyses of candidates with subsequent observations from the ground. 

As the population of planets with known masses and radii increased, a deficiency of sub-Jovian planets with P < 10 days became evident. This phenomenon, known as the Neptunian or sub-Jovian desert \citep{2016A&A...589A..75M}, is seen as approximately triangular regions in the radius--period and mass--period parameter spaces (see Fig.~\ref{fig:desert}) and is not due to observational biases. The study of planets within the desert or at its boundaries can offer clues as to the different formation and evolution mechanisms at play.

In addition to allowing mass and radius determinations, each transiting planet found around a bright star opens up numerous characterization possibilities. One can sometimes obtain emission and transmission spectra \citep{2002ApJ...568..377C, 2005sptz.prop20523C}, measure the spin-orbit obliquity \citep{2018haex.bookE...2T, 2010ApJ...724.1108J}, and even probe the atmospheric structure by analyzing phase curves \citep{2018haex.bookE.116P}. 

In this work we performed a detailed analysis of two stars from the TESS input catalog that had been identified as planet host candidates or TESS objects of interest (TOIs), namely TOI-1199 (TIC 99869022) and TOI-1273 (TIC 445859771). We combined photometric data from TESS, KeplerCam, and MuSCAT2 (see Sect.~\ref{subsec:lc_followup}) with precise RV measurements from the SOPHIE spectrograph (see Sect.~\ref{subsec:sophie}) to determine the presence of the planets and characterize their properties, including measurements of their bulk densities. 

The following section is a description of the observations, in Sect.~\ref{sec:analysis} we present the analysis of the stellar and planetary parameters, Sect.~\ref{sec:discussion} provides a discussion of the results, and in Sect.~\ref{sec:conclusion} we report the conclusions of our work.
\section{Observations}
\label{sec:observations}
\begin{figure*}[h!]
\resizebox{\hsize}{!}
        {\includegraphics{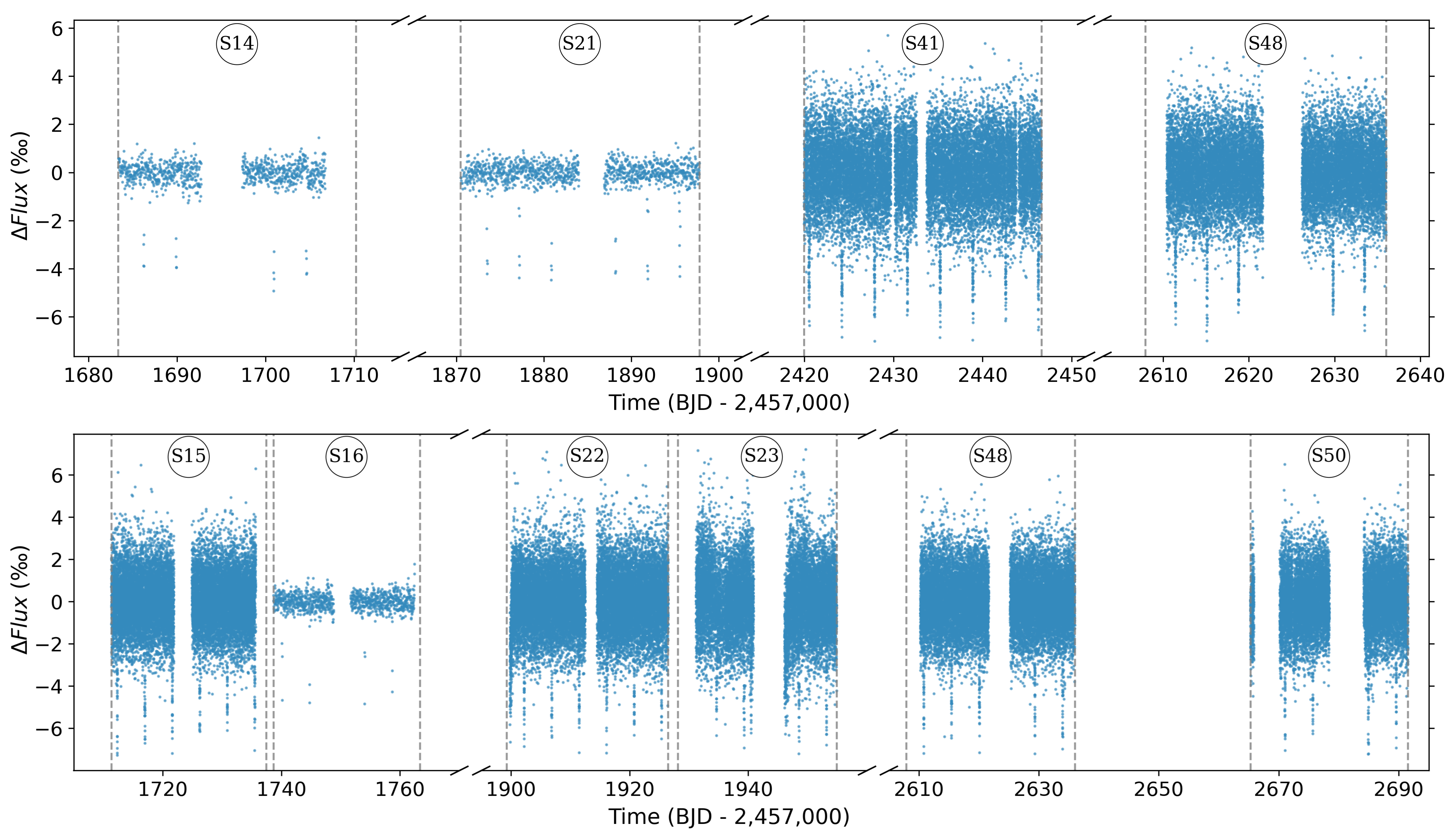}}
        \caption{TESS photometric data used for TOI-1199 (top) and TOI-1273 (bottom). Dashed lines mark the different TESS sectors, and each sector number is labeled. Transit features are easily detected in both panels.}
     \label{fig:tessdata}
\end{figure*}
\subsection{TESS photometry}
\label{subsec:tessphotometry}
TOI-1199 was observed with TESS with 30-minute cadence during sectors 14 and 21 and with 2-minute cadence during sectors 41 and 48, totaling around 111 days between 2020 and 2022 and obtaining 23 full transits. TOI-1273 was monitored with 2-minute cadences in sectors 15, 22, 23, 48, and 50, and with 30-minute cadence in sector 16, summing up to around 168 days between August 2019 and April 2022 with 32 transit events, two of which were not complete because they fell at the beginning or end of a sector.  

The transit signature of TOI-1199.01 was detected by the Quick Look Pipeline \citep[QLP;][]{2020RNAAS...4..204H, 2020RNAAS...4..206H}, which performed a suite of diagnostic tests favoring the planetary nature of the signal and fitted a limb-darkened transit model. The TESS Science Office (TSO) reviewed this information and issued an alert on 27 August 2019 \citep{2021ApJS..254...39G}. The TESS Science Processing Operations Center (SPOC) pipeline subsequently detected the same signature in searches of 2-min data in sectors 41 and 48 with a noise-compensating matched filter \citep{2002ApJ...575..493J, 2010SPIE.7740E..0DJ, 2020TPSkdph}, and the signature was fitted with a limb-darkened transit model \citep{Li:DVmodelFit2019} and passed all the diagnostic tests performed \citep{Twicken:DVdiagnostics2018}, including the difference image centroiding test, which located TOI-1199 to within 1.2\,$\pm$\,2.8 arcsec.
The transit signature of TOI-1273.01 was detected by both the QLP and SPOC pipelines in transit searches of sector 15 and alerted by the TSO on 17 October 2019. The difference image centroiding test performed by the SPOC for sector 15 located the host star within 0.5880\,$\pm$\,2.5565 arcsec of the source of the transit, and subsequently to within  0.3240\,$\pm$\,2.4923 arcsec  based on an analysis of sectors 15, 22, 23, 48 \& 50.

\begin{figure*}[h!]
\centering
\includegraphics[width=\hsize]{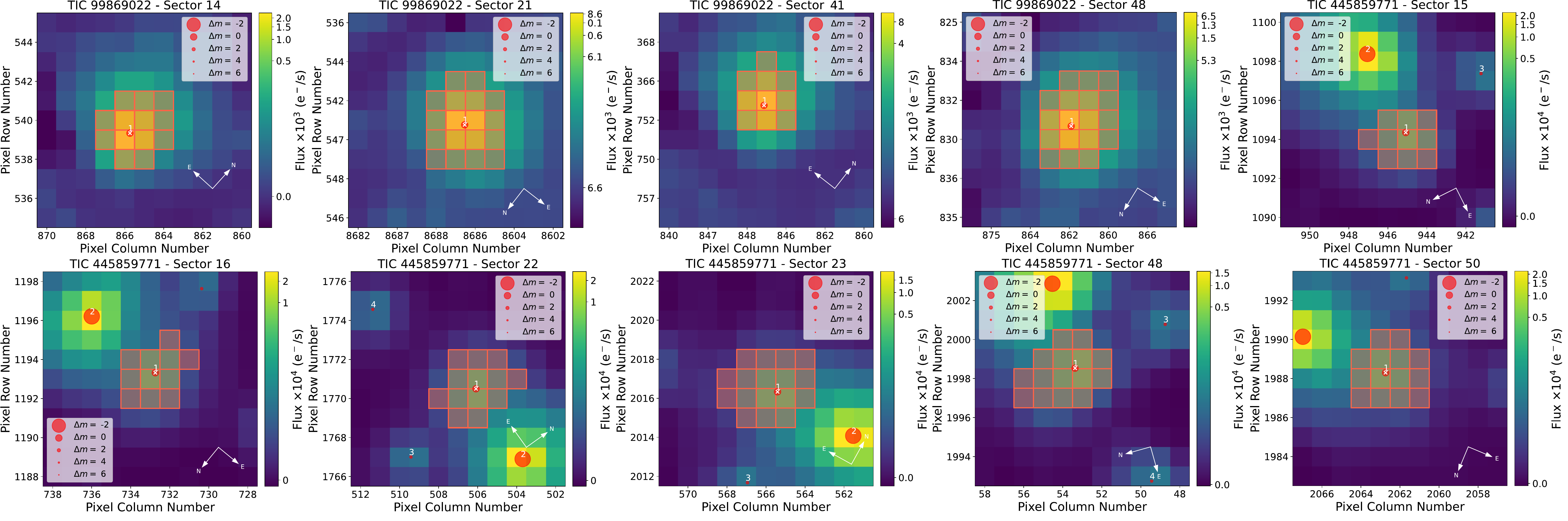}
\caption{TESS target pixel file images of all sectors used in our analysis. The four images in the first row from left to right correspond to sectors 14, 21, 41, and 48 for TOI-1199. The last image of the first row and the five from the second row correspond to sectors 15, 16, 22, 23, 48, and 50 for TOI-1273. Colors show the electron counts. Pixels colored in red were used for the simple aperture photometry. The target star and nearby sources at their \textit{Gaia} DR3 positions \citep{2023A&A...674A...1G}  are marked with numbered red circles; number 1 is the target star. The size of the circles codes their relative magnitude with respect to the target star.}
\label{fig:tpf}
\end{figure*}

We downloaded from the Mikulski Archive for Space Telescopes (MAST\footnote{\url{https://mast.stsci.edu/}}) the 2-minute cadence light curves computed by the TESS SPOC pipeline \citep{2016SPIE.9913E..3EJ} and the 30-minute cadence light curves, which have the same preprocessing but are derived from TESS full frame images by \citet{2020RNAAS...4..201C}. We then used the \texttt{Lightkurve} python package \citep{2018ascl.soft12013L} to remove all NaNs and 5$\sigma$ outliers. We used the column \texttt{pdcsap\_flux} as the flux, which is the light curve from the Presearch Data Conditioning Simple Aperture Photometry (PDCSAP) and is corrected for crowding contamination, instrumental trends and noise \citep{Stumpe2012, Stumpe2014, 2012PASP..124.1000S}. The data for both targets are presented in Fig.~\ref{fig:tessdata}, where the flux is shown in parts per thousand (\textperthousand) and the transits can be seen at plain sight. The depths and durations of the transits reported by TESS-SPOC are 0.43\% and  2.1$\,\pm\,$0.2\,h for TOI-1199.01 and 0.48\% and 1.4$\,\pm\,$0.2\,h for TOI-1273.01. Contamination ratios of 0.047\% for TOI-1199 and 0.30\% for TOI-1273 are also reported, where the larger value for TOI-1273 is most likely due to a close source as can be seen in the TESS target pixel file images shown in Fig.~\ref{fig:tpf}, which were made using \texttt{tpfplotter}\footnote{\url{https://github.com/jlillo/tpfplotter}} \citep{2020A&A...635A.128A}. 

Given the good quality of the PDCSAP light curves, no further detrending or contamination correction was deemed necessary and we did not include a dilution factor in the models. We ran a box least-squares analysis \citep{2016ascl.soft07008K} of the TESS time series, which recovered the periodic signals at 3.67\,d for TOI-1199.01 and 4.63\,d for TOI-1273.01. The phase-folded light curve of TOI-1273.01 reveals a well-defined V-shape transit (see Fig.~\ref{fig:fit1273}), meaning that if this is product of a planetary transit, the transit must be close to grazing (i.e., the transit impact parameter is near 1).

\subsection{Light curve follow-up}
\label{subsec:lc_followup}
The TESS image scale is $\sim$\,21\arcsec\,per pixel and photometric apertures typically extend out to roughly 1 arcmin, generally causing multiple stars to blend in the TESS aperture. While the SPOC pipeline tests already constrains the location of the source of the transit signature to within a pixel (see Sect.~\ref{subsec:tessphotometry}), to further confirm the true source of the TOI detections, we conducted ground-based photometric follow-up observations of the field around TOI-1199 and TOI-1273 as part of the TFOP\footnote{\href{https://tess.mit.edu/followup}{https://tess.mit.edu/followup}} Sub Group 1 \citep[][]{collins:2019}. We used the {\tt TESS Transit Finder}, which is a customized version of the {\tt Tapir} software package \citep{Jensen:2013}, to schedule our transit observations. The images were calibrated and differential photometric data were extracted using {\tt AstroImageJ} \citep{Collins:2017}, except as noted below. All light curve data are available on the {\tt EXOFOP-TESS} website for both TOI-1199.01\footnote{\href{https://exofop.ipac.caltech.edu/tess/target.php?id=99869022}{https://exofop.ipac.caltech.edu/tess/target.php?id=99869022}} and TOI-1273.01\footnote{\href{https://exofop.ipac.caltech.edu/tess/target.php?id=445859771}{https://exofop.ipac.caltech.edu/tess/target.php?id=445859771}}. A summary of all the photometry data is presented in Table~\ref{tab:photometry}. 

\begin{table*}
\caption{Summary of photometric observations.}             
\centering          
\begin{tabular}{ l c c c c}     
\hline\hline       
Date & Facility \& Instrument & Transit coverage & Filter & Used in the model? \\
\hline
\multicolumn{5}{c}{TOI-1199} \\
18 Jul 2019 & TESS 30-min & Full (4) & TESS & Yes \\
21 Jan 2020 & TESS 30-min & Full (6) & TESS & Yes \\
24 Jul 2021 & TESS 2-min & Full (8) & TESS & Yes \\
28 Jan 2022 & TESS 2-min & Full (5) & TESS & Yes \\
13 Feb 2021 &  \textit{Fred Lawrence Whipple} Observatory - KeplerCam & Full (1) & \textit{B}, \textit{z'} & Yes \\
13 Apr 2021 &  \textit{Fred Lawrence Whipple} Observatory - KeplerCam & Full (1) & \textit{B}, \textit{z'} & Yes \\
27 Apr 2021 & \textit{Carlos Sanchez} Telescope - MuSCAT2 & Egress & \textit{g'}, \textit{r'}, \textit{i'}, \textit{z$_s'$} & No \\
4 Feb 2020 & CROW Observatory & Full (1) & g' & No \\
4 Feb 2020 & Wild Boar Remote Observatory  & Full (1) & \textit{R$_c$} & No \\
9 Dec 2020 & Waffelow Creek Observatory & Full (1) & \textit{g'} & No \\
\hline
\multicolumn{5}{c}{TOI-1273} \\
15 Aug 2019 & TESS 2-min & Full (6) & TESS & Yes \\
12 Sep 2019 & TESS 30-min & Full (4) & TESS & Yes \\
19 Feb 2020 & TESS 2-min & Full (6) & TESS & Yes \\
19 Mar 2020 & TESS 2-min & Full (5) & TESS & Yes \\
28 Jan 2022 & TESS 2-min & Full (5) & TESS & Yes \\
26 Mar 2022 & TESS 2-min & Full (4) & TESS & Yes \\
13 Feb 2020 &  \textit{Fred Lawrence Whipple} Observatory - KeplerCam & Full (1) & \textit{B} & Yes \\
18 Feb 2020 & \textit{Carlos Sanchez }Telescope - MuSCAT2 & Ingress & \textit{g'}, \textit{r'}, \textit{i'}, \textit{z$_s'$} & Yes \\
21 Jan 2020 & Acton Sky Portal & Full (1) & \textit{r'} & No \\
17 Feb 2020 & Catania Astrophysical Observatory & Partial & \textit{B} & No \\
\hline                  
\end{tabular}
\tablefoot{TESS dates correspond to the start date of the sector. When full transits are observed, the number of transits is indicated in parenthesis.}
\label{tab:photometry}
\end{table*}

\subsubsection {Citizen scientist observations}

We initially observed TOI-1199 using citizen scientists observatories. On 4 February 2020 we simultaneously observed a transit of TOI-1199.01 from the 0.36\,m telescope at CROW observatory (Portalegre, Portugal) in Sloan \textit{g'} band and the 0.24\,m telescope at Wild Boar Remote Observatory (San Casciano in val di pesa, Firenze, Italy) in \textit{Rc} band and detected roughly 0.4\% deep events on target using photometric apertures that exclude flux from the nearest known \textit{Gaia} Data Release 3 (DR3) star $\sim$16$\arcsec$ north of TOI-1199. We observed a second transit epoch with the 0.36\,m telescope at Waffelow Creek Observatory (Nacogdoches, TX, USA) on 9 December 2020 in Sloan \textit{g'} band and again confirmed the event on target relative to known \textit{Gaia} DR3 stars. We adjusted the follow-up ephemeris from these light curves to a slightly longer period, which allowed us to predict  the transit timings for our later follow-up observations with a higher precision. 

We observed a transit of TOI-1273.01 from the 0.36\,m telescope at Acton Sky Portal (Acton, MA, USA) on 21 January 2020 in Sloan \textit{r'} band and detected a roughly 0.5-0.6\% deep event on target relative to the nearest known \textit{Gaia} DR3 star $\sim$13$\arcsec$ northeast of TOI-1273. While this transit observations were important in the follow-up stage of the candidates, they were not included in the model fit. However, we show the best-fit model overplotted to the phased light curves in Appendix~\ref{apx:extralcs}. 

\subsubsection{MuSCAT2}

A transit event of TOI-1199.01 was observed on 27 April 2020 with MuSCAT2 \citep{Narita2019} mounted on the 1.5\,m \textit{Carlos Sanchez }Telescope at Teide Observatory, Spain. MuSCAT2 is a multi-band imager equipped with four 1024$\times$1024 pixels CCDs with a field of view of $7\farcm4\times7\farcm4$. It can obtain near-simultaneous images in \textit{g'}, \textit{r'}, \textit{i'}, and \textit{z$_s'$} bands and it was designed for transiting planets follow-up observations.

During the observations the telescope was slightly defocused to avoid the saturation of the target and clouds were present for roughly half the transit. The exposure times were set to 15, 10, 15, and 15\,s in \textit{g'}, \textit{r'}, \textit{i}', and \textit{z$_s'$}, respectively. The raw data were reduced by the MuSCAT2 pipeline \citep{Parviainen2019}; the pipeline performs dark and flat field calibrations, aperture photometry and instrumental systematics correction. The single transit event was interrupted by clouds on the first half of the transit. We attempted to use the second half for the modeling, but there was additional systematic noise in the data. When we removed the transit from the fit, there was no significant impact on the derived parameters, so we decided not to include it in the final model. The data are shown in Appendix~\ref{apx:extralcs} with the best-fit model overplotted. TOI-1273 was observed on the night of 17 February 2020 with MuSCAT2 in \textit{g'}, \textit{r'}, \textit{i'}, and \textit{z$_s'$} bands. The data were acquired with the telescope slightly defocused and the exposure times were set to 8\,s for all bands. The raw data were also calibrated and reduced with the MuSCAT2 pipeline \citep{Parviainen2019}. Again, the transit captured was affected by clouds, this time on the second half. We used the first half of the transit on the four bands in the model. The data are shown in Fig.~\ref{fig:fit1273} along the best-fit model.

\subsubsection{KeplerCam}

We observed two full transits of TOI-1199.01 on 13 February 2021 and 13 April 2021 with the KeplerCam instrument on the 1.2\,m telescope at the \textit{Fred Lawrence Whipple} Observatory (FLWO) using alternating \textit{B} band and Sloan \textit{z'} band filters, resulting in four light curves. The 4096$\times$4096 Fairchild CCD 486 detector has an image scale of 0\farcs336 per pixel, resulting in a $23\farcm1\times23\farcm1$ field of view. We also observed one full transit of TOI-1273.01 on 13 February 2020 with KeplerCam in \textit{B} band. The KeplerCam light curves of both targets were used in the models and are shown on Fig.~\ref{fig:fit1199} and Fig.~\ref{fig:fit1273}.

\subsubsection{Catania Astrophysical Observatory}

We observed one almost full transit of TOI-1273.01 (a part of the egress is missing) on 17 February 2020 in \textit{B} band from the 0.91\,m telescope at Catania Astrophysical Observatory (Catania, Italy) and detected a roughly 0.5-0.6\,\% event on target. The custom built 1024$\times$1024 detector uses a KAF1001E CCD with an image scale of $0\farcs66$ per pixel, resulting in a $11\farcm2\times11\farcm2$ field of view. These data were not used in the model and is shown in Appendix~\ref{apx:extralcs}.

\subsection{SOPHIE spectroscopy}
\label{subsec:sophie}
Radial velocity measurements for the two targets were obtained from high-resolution spectroscopy with SOPHIE\footnote{\url{http://www.obs-hp.fr/guide/sophie/sophie-eng.shtml}}, the fiber-fed echelle spectrograph mounted on the $1.93~$m telescope at the Haute-Provence Observatory in France \citep{2008SPIE.7014E..0JP, 2013A&A...549A..49B}. The spectrograph is fed from the Cassegrain focus through either one of two separate optical fiber sets, yielding two different spectral resolutions (high-efficiency and high-resolution modes). The spectrograph covers the wavelength range 3872-6943\,\r{A}. Using the high-resolution mode (R\,=\,75000) we obtained 60 RVs for TOI-1199 between December 2019 and June 2022 with a mean uncertainty of 4.5\,\ms\,, mean signal-to-noise ratio (S/N) of $\sim$30 and mean exposition time of 1333\,s. We also obtained 60 RVs for TOI-1273, which have a mean uncertainty of 5.4\,\ms\,, mean S/N\,$\sim$32, mean exposition time of 1500\,s and were observed between February 2020 and July 2022. 

The RVs were derived through the SOPHIE pipeline \citep{2009A&A...505..853B},
making cross-correlations with numerical masks. We used the optimized procedures presented by \citet{2022A&A...658A.176H} and \citet{2024A&A...681A..55H}. This includes in particular: (1) CCD charge transfer inefficiency correction \citep{2009EAS....37..247B}; (2) correction for the moonlight contamination using the simultaneous sky spectrum obtained from the second SOPHIE fiber aperture  \citep{2008MNRAS.385.1576P, 2008A&A...488..763H}; (3) RV constant master correction for instrumental long-term drifts \citep{2015A&A...581A..38C}; and (4) correction of the instrumental short-term drifts thanks to the frequently measured drifts interpolated at the precise time of each observation. 

\begin{figure}[h!]
\centering
\includegraphics[width=\hsize]{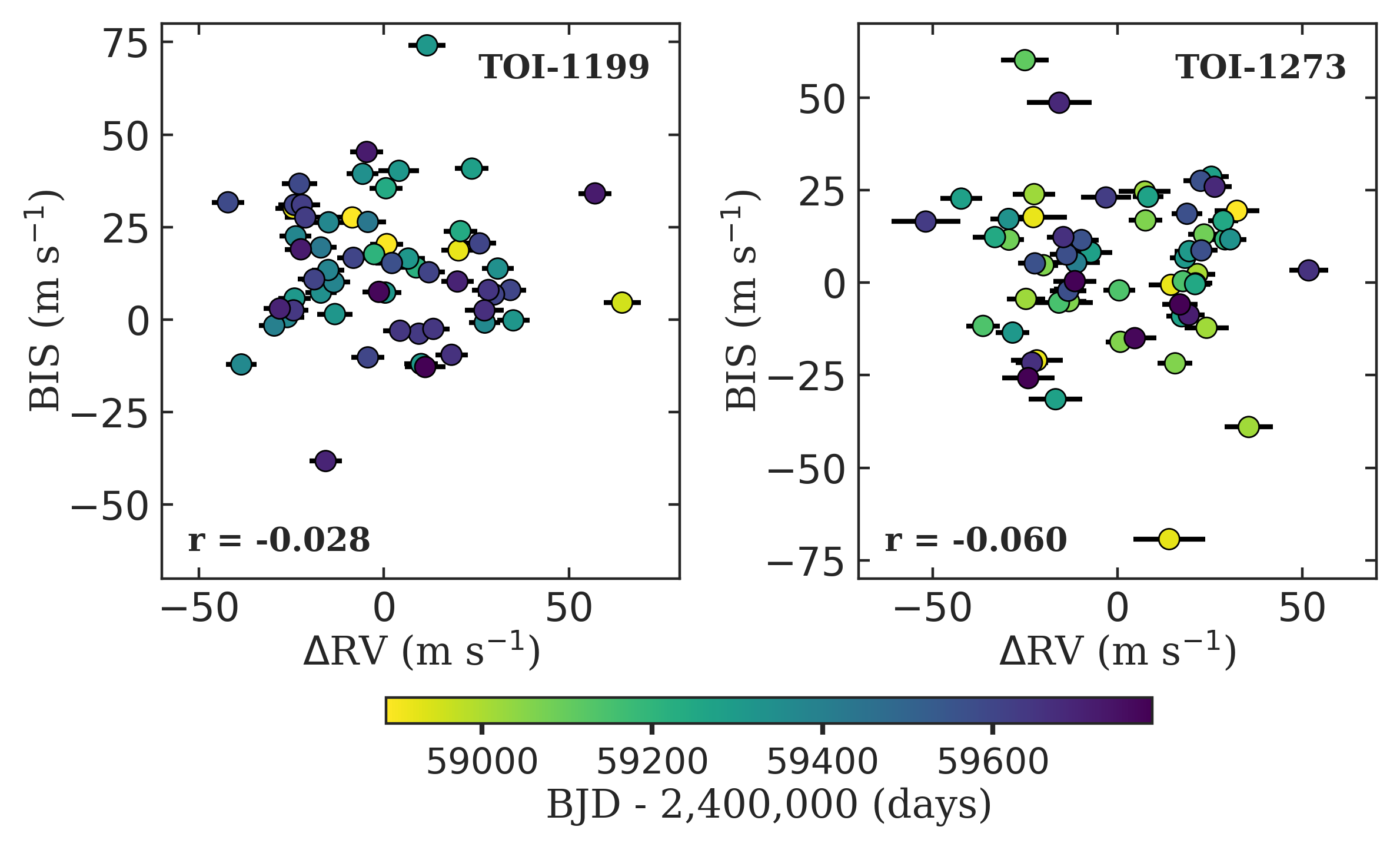}
\caption{SOPHIE RVs plotted against the BISs of the cross-correlation function. The barycentric Julian date (BJD) of each observation is color coded. Pearson's $r$ correlation coefficient  cases indicates no significant linear correlation between the RVs and BISs in either case. TOI-1199 (left panel) and TOI-1273 (right panel).}
\label{fig:bisrv}
\end{figure}

\begin{figure*}[h!]
\centering
\includegraphics[width=\hsize]{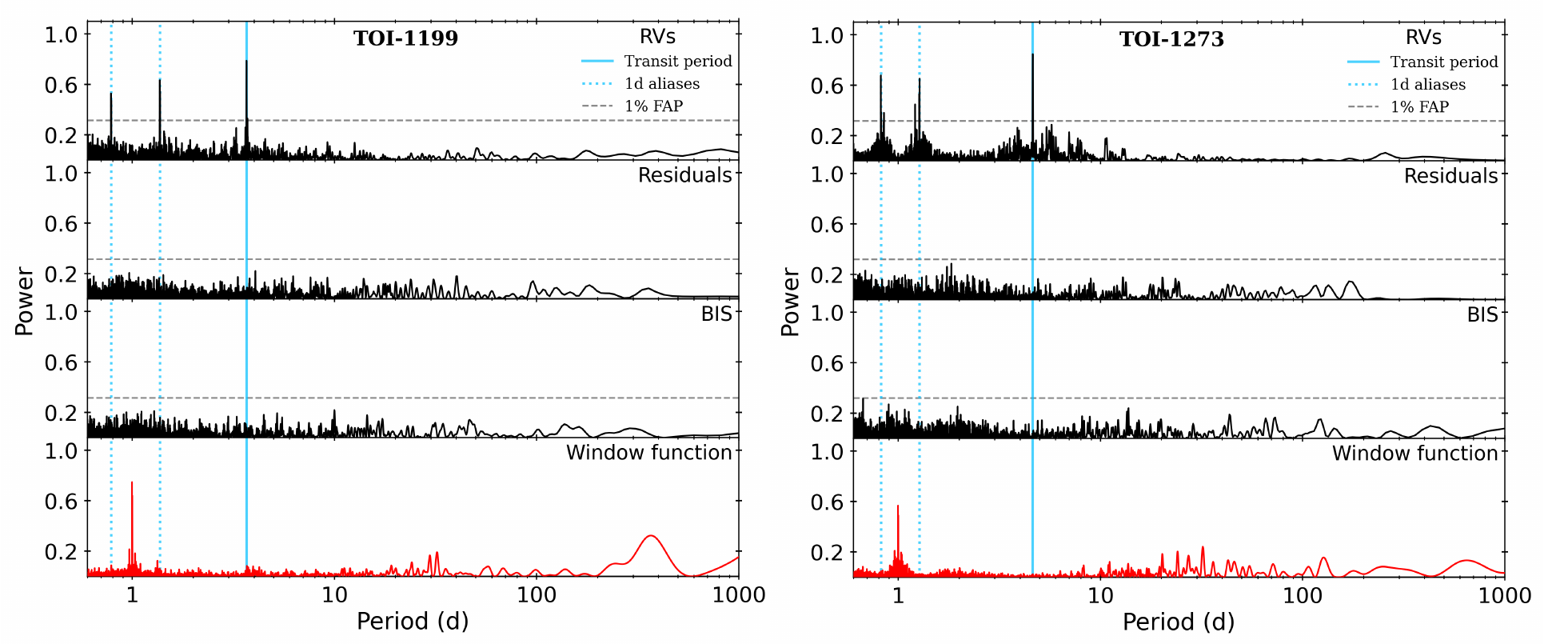}
\caption{GLS periodograms of the SOPHIE RV time series, the residuals of the best-fit Keplerian model (from Sect. \ref{planetarysubsection}), the BIS, and the window function. For TOI-1199 (\textit{left panel}) and TOI-1273 (\textit{right panel}), the image shows the computed power spectrum of the GLS. In the RVs, the highest peak matches the transit period of the planet candidates, but the same signal is not present in the residuals, the BIS, or the window function. The analytical false alarm probability threshold of 1\% is shown.}
\label{fig:periodograms}
\end{figure*}

We excluded the 15 bluest spectral orders from the cross-correlations due to their low S/N. We tried different masks characteristic of various stellar types, which all produced results in agreement. This agreement favors the planetary scenario, whereas a transit implying blended stars of different spectral types might
produce RV semi-amplitudes varying with the stellar mask. The bisector spans (BISs) of each cross-correlation function were also computed following \citet{2001A&A...379..279Q}. They show no significant variations nor correlation with the RVs (see Fig.~\ref{fig:bisrv}), also arguing in favor of the planetary scenario for the transit events. Figure \ref{fig:periodograms} shows the generalized Lomb-Scargle\footnote{We used the \texttt{astropy} implementation of this method \citep{2022ApJ...935..167A}.} \citep[GLS;][]{2009A&A...496..577Z} periodograms of both targets. For the two systems, the RV shows variations in agreement with the periods and phases derived from the photometry (Sect.~\ref{subsec:tessphotometry} and \ref{subsec:lc_followup}). We see no further significant signals in the residuals after removing the best-fit model or in the BISs. From these results, we validate TOI-1199.01 and TOI-1273.01 as planetary transits, and forward refer to the planets as TOI-1199\:b and TOI-1273\:b. The complete RV and BIS time series for both targets are displayed in Appendix~\ref{apx:rvs}. 

\subsection{High-resolution imaging}
If an exoplanet host star has a spatially close companion, that companion (bound or line of sight) can create a false-positive transit signal if it is, for example, an eclipsing binary. Moreover, ``third-light” flux from the close companion star can lead to an underestimated planetary radius if not accounted for in the transit model \citep{2015ApJ...805...16C}, yield incorrect planet and stellar parameters \citep{2017AJ....154...66F, 2020ApJ...898...47F}, and cause non-detections of small planets residing with the same exoplanetary system \citep{2021AJ....162...75L}. Additionally, the discovery of close, bound companion stars, which exist in nearly one-half of FGK-type stars \citep{2018AJ....156...31M} provides crucial information toward our understanding of exoplanetary formation, dynamics, and evolution \citep{2021FrASS...8...10H}. Thus, to search for close-in bound companions unresolved in photometric observations (Sect.~\ref{subsec:tessphotometry} and \ref{subsec:lc_followup}) and undetected from SOPHIE stellar mask and bisector studies (Sect.~\ref{subsec:sophie}), we obtained high-resolution imaging speckle observations of TOI-1199 and TOI-1273.

\begin{figure*}[h!]
\centering
\includegraphics[width=1\hsize]{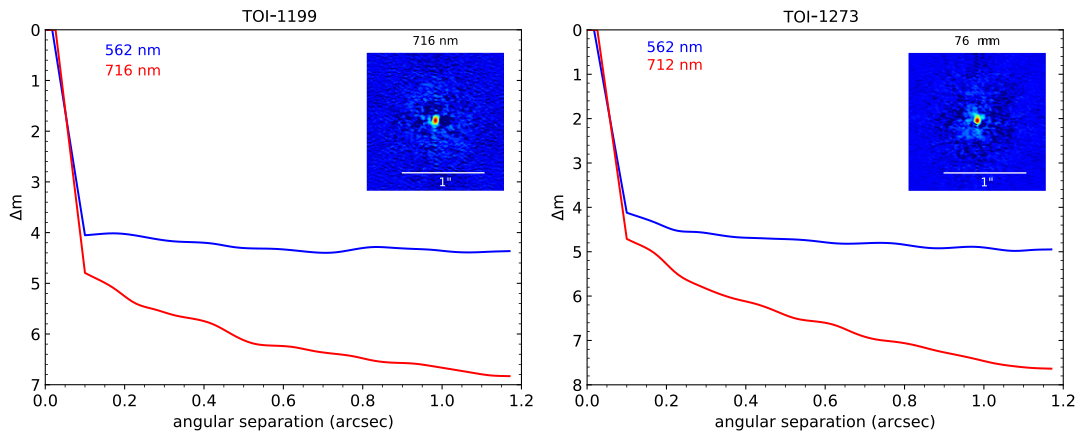}
\caption{Gemini speckle observations of TOI-1199 (\textit{left}) and TOI-1273 (\textit{right}). The contrast curves in the 562\,nm (\textit{red}) and 832\,nm (\textit{blue}) bands are shown along with the reconstructed 832\,nm image. The data show that there are no close-in companions brighter than 5-7 magnitudes below the target stars magnitude from 0.1$\arcsec$ to 1.2$\arcsec$.}
\label{fig:speckle}
\end{figure*}

TOI-1199 and TOI-1273 were observed on 15 and 17 February 2020 (respectively) using the ‘Alopeke speckle instrument on the Gemini North 8 m telescope\footnote{\url{https://www.gemini.edu/sciops/instruments/alopeke-zorro/}} \citep{2021FrASS...8..138S}.  ‘Alopeke provides simultaneous speckle imaging in two bands (562\,nm and 832\,nm) with output data products including a reconstructed image with robust contrast limits on companion detections \citep[e.g.,][]{2022FrASS...9.1163H}. Four sets of 1000$\,\times\,$0.06\,s exposures were collected for each target and subjected to Fourier analysis in our standard reduction pipeline \citep[see][]{2011AJ....142...19H}. Figure~\ref{fig:speckle} shows our final contrast curves and the 832 nm reconstructed speckle images. We find that both TOI-1199 and TOI-273 are single stars revealing no nearby companion brighter than 5-7 magnitudes below that of the target star from 0.1" out to 1.2”. At the distance of TOI-1199 ($d$\,=\,247\,pc) and TOI-1273 ($d$\,=\,176\,pc) these angular limits correspond to spatial limits of 25-298\,AU and 18-212\,AU, respectively.
\section{Analysis}
\label{sec:analysis}
\begin{table*}
\caption{Host star parameters.}       
\label{tab:stellarparam}
\centering
\begin{tabular}{l c c c c c}     
\hline\hline       
 &  \multicolumn{2}{c}{TOI-1199} & \multicolumn{2}{c}{TOI-1273} &  \\
\hline
 Parameter & \multicolumn{4}{c}{} & Ref. \\
\hline  
\multicolumn{6}{c}{Designations} \\
TIC & \multicolumn{2}{c}{99869022} & \multicolumn{2}{c}{445859771} & (1) \\
{2MASS} & \multicolumn{2}{c}{J11073136+6121096} & \multicolumn{2}{c}{J14162891+5823255} & (2) \\
\textit{Gaia} DR3 & \multicolumn{2}{c}{861975270310252416} & \multicolumn{2}{c}{1611685004650291840} & (3) \\
\hline  
\multicolumn{6}{c}{Astrometry} \\
RA (J2016.0) & \multicolumn{2}{c}{11:07:31.35} & \multicolumn{2}{c}{14:16:29.01} & (3) \\
Dec. (J2016.0) & \multicolumn{2}{c}{+61:21:9.24} & \multicolumn{2}{c}{+58:23:24.89} & (3)\\
$\mu \cos{\delta}$ (mas) & \multicolumn{2}{c}{$ -3.963\pm0.013$} & \multicolumn{2}{c}{$37.200\pm0.012$} & (3)\\
$\pi$ (mas) & \multicolumn{2}{c}{$4.041\pm0.016$} & \multicolumn{2}{c}{$5.659\pm0.010$} & (3)\\
Distance (pc) & \multicolumn{2}{c}{$247.0^{+0.8}_{-0.7}$} & \multicolumn{2}{c}{$176.0^{+0.4}_{-0.4}$}& (4)\\
\hline 
\multicolumn{6}{c}{Photometry}  \\
Near-UV & \multicolumn{2}{c}{--} & \multicolumn{2}{c}{$16.756\pm0.019$} & (5) \\
$B$ & \multicolumn{2}{c}{$11.675\pm0.099$} & \multicolumn{2}{c}{$11.798\pm0.055$} & (6) \\
$V$ & \multicolumn{2}{c}{$11.069\pm0.069$} & \multicolumn{2}{c}{$11.011\pm0.063$} & (6) \\
$g'$ & \multicolumn{2}{c}{$11.272\pm0.072$} & \multicolumn{2}{c}{$11.385\pm0.019$} & (6) \\
$r'$ & \multicolumn{2}{c}{$10.873\pm0.070$} & \multicolumn{2}{c}{$10.80\pm0.10  $} & (6) \\
$i'$ & \multicolumn{2}{c}{$10.699\pm0.081$} & \multicolumn{2}{c}{$10.60\pm0.11$} & (6) \\
$G$ & \multicolumn{2}{c}{$10.8858\pm0.0001$} & \multicolumn{2}{c}{$10.8656\pm0.0001$} & (3) \\
$G_\mathrm{BP}$ & \multicolumn{2}{c}{$11.2413\pm0.0004$} & \multicolumn{2}{c}{$11.2050\pm0.0004$} & (3) \\
$G_\mathrm{RP}$ & \multicolumn{2}{c}{$10.3684\pm0.0003$} & \multicolumn{2}{c}{$10.3603\pm0.0003$} & (3) \\
$J$ & \multicolumn{2}{c}{$9.830\pm0.021$} & \multicolumn{2}{c}{$9.814\pm0.022$} & (2) \\
$H$ & \multicolumn{2}{c}{$9.542\pm0.017$} & \multicolumn{2}{c}{$9.515\pm0.028$} & (2) \\
$K_S$ & \multicolumn{2}{c}{$9.466\pm0.015$} & \multicolumn{2}{c}{$9.417\pm0.024$} & (2) \\
$W_1$ & \multicolumn{2}{c}{$9.410\pm0.023$} & \multicolumn{2}{c}{$9.385\pm0.022$} & (7) \\
$W_2$ & \multicolumn{2}{c}{$9.464\pm0.019$} & \multicolumn{2}{c}{$9.430\pm0.019$} & (7) \\
$W_3$ & \multicolumn{2}{c}{$9.398\pm0.032$} & \multicolumn{2}{c}{$9.375\pm0.031$} & (7) \\
\hline 
 \multicolumn{6}{c}{Bulk Properties}  \\
   $T_\mathrm{eff}\,$(K) & \multicolumn{2}{c}{$5710\pm40$} & \multicolumn{2}{c}{$5690\pm40$} & (8) \\
   log\,$g$ & \multicolumn{2}{c}{$4.19\pm0.03$} &  \multicolumn{2}{c}{$4.37\pm0.03$} & (8) \\
   $V_\mathrm{turbulence}$\,(\ms) & \multicolumn{2}{c}{$0.91\pm0.04$} & \multicolumn{2}{c}{$0.81 \pm 0.03$} & (8) \\ 
   $[Fe/H]~$(dex) & \multicolumn{2}{c}{$0.44 \pm 0.04$} & \multicolumn{2}{c}{$0.07 \pm 0.04$} & (8) \\
   $v\,\sin{i}~$(\kms) & \multicolumn{2}{c}{$3.4\pm0.4$} & \multicolumn{2}{c}{$2.8\pm0.4$}  & (8) \\
   log $R'_\text{HK}$ & \multicolumn{2}{c}{$-5.1 \pm 0.2$} & \multicolumn{2}{c}{$-5.0 \pm 0.2$} & (8) \\
   $A_V$  & \multicolumn{2}{c}{$0.02 \pm 0.02$} &   \multicolumn{2}{c}{$0.01 \pm 0.01$} & (8) \\
   $F_{\rm bol}$~( $10^{-9}$~erg~s$^{-1}$~cm$^{-2}$)  & \multicolumn{2}{c}{$1.0553 \pm 0.0017$} &   \multicolumn{2}{c}{$1.097 \pm 0.013$} & (8) \\
   $L_{\rm bol}$~(L$_{\rm bol}$) & \multicolumn{2}{c}{$2.015 \pm 0.008$} & \multicolumn{2}{c}{$1.07 \pm 0.01$}  & (8) \\
   Radius (\RS) & \multicolumn{2}{c}{$1.45 \pm 0.03$} & \multicolumn{2}{c}{$1.06 \pm 0.02$} & (8) \\
   Mass (\msol) & \multicolumn{2}{c}{$1.23 \pm 0.07$} &  \multicolumn{2}{c}{$1.06 \pm 0.06$} & (8) \\
   Age (Gyr) & \multicolumn{2}{c}{$4.2 \pm 0.2$} &  \multicolumn{2}{c}{$3.1 \pm 1.6$} & (8) \\
   $P_{\rm rot}$~(d) & \multicolumn{2}{c}{$23 \pm 5$} &  \multicolumn{2}{c}{$42 \pm 11$} & (8) \\
   $P_{\rm rot}/\sin i$~(d) & \multicolumn{2}{c}{$21.9 \pm 0.6$} &  \multicolumn{2}{c}{$20.7 \pm 2.4$} & (8) \\
\hline                  
\end{tabular}
\tablefoot{(1) TIC \citep{2019AJ....158..138S}. (2) \textit{2MASS} \citep{2003yCat.2246....0C}. (3) \textit{Gaia} DR3 \citep{2023A&A...674A...1G}. (4) \cite{2021AJ....161..147B}. (5) \textit{GALEX} \citep{2011Ap&SS.335..161B}. (6) \textit{APASS} \citep{2015AAS...22533616H}. (7) \textit{WISE} \citep{2014yCat.2328....0C}. (8) This work (Sect.~\ref{subsec:stellarparam}).}
\end{table*}
\subsection{Stellar parameters} \label{subsec:stellarparam}
TOI-1199 and TOI-1273 are both G-type dwarfs, with V magnitudes of $\sim$11 that reside in the vicinity of the Sun, with distances 247.0$^{+0.8}_{-0.7}$\,pc and 176.0$^{+0.4}_{-0.4}$\,pc, respectively, as reported by \citet{2021AJ....161..147B} from \textit{Gaia} parallaxes.

From the combined SOPHIE spectra unpolluted by moonlight the stellar atmospheric parameters ($T_{\mathrm{eff}}$, $\log g$, $V_\mathrm{turbulence}$, [Fe/H]) were derived using the ARES+MOOG methodology described in \citet{2021A&A...656A..53S}, \citet{2014dapb.book..297S}, and  \citet{2013A&A...556A.150S}. We use the latest version of ARES\footnote{The last version, ARES v2, can be downloaded at \url{https://github.com/sousasag/ARES}} \citep{2007A&A...469..783S, 2015A&A...577A..67S} to measure the equivalent widths of selected iron lines on the combined spectrum of TOI-1199 and TOI-1273. The list of iron lines is the same as the one presented in \citet{2008A&A...487..373S}. A minimization process is used to find the ionization and excitation equilibrium and converge to the best set of spectroscopic parameters. This process makes use of a grid of Kurucz model atmospheres \citep{1993sssp.book.....K} and the radiative transfer code MOOG \citep{1973PhDT.......180S}. This procedure leads to the following atmospheric parameters: $T_{\mathrm{eff}}=5700\pm66$\,K, $\log g=4.15\pm0.12$, $V_\mathrm{turbulence}=0.91\pm0.04$\,\ms, $[Fe/H]=0.42\pm0.05$\,dex, and $v~\sin{i}=3.3\pm1.0$\,\kms\, for TOI-1199 and $T_{\mathrm{eff}}=5700\pm70$\,K, $\log g=4.34\pm0.10$, $V_\mathrm{turbulence}=0.81\pm0.03$\,\ms, $[Fe/H]=0.07\pm0.04$\,dex, and $v~\sin{i}=2.2\pm1.0$\,\kms\,for TOI-1273. We also derived a more accurate trigonometric surface gravity using recent \textit{Gaia} data following the same procedure as described in \citet{2021A&A...656A..53S}, leading to $\log g=4.19\pm0.03$ and $\log g=4.38\pm0.03$ for TOI-1199 and TOI-1273, respectively. The log $R'_\text{HK}$ index were computed following \citet{2010A&A...523A..88B}, indicating low activity for both stars, with values of $-5.1\pm0.2$ and $-5.0\pm0.2$ for TOI-1199 and TOI-1273, respectively.

As an independent analysis we also derived stellar parameters using spectra obtained with the Tillinghast Reflector Echelle Spectrograph \citep[TRES;][]{gaborthesis} on the 1.5\,m Tillinghast Reflector at FLWO in Arizona, USA. TRES is an optical (390-910\,nm) fiber-fed echelle spectrograph with a resolving power of $\sim$44,000. The spectra were extracted using the standard pipeline as described in \citet{2010ApJ...720.1118B} and stellar parameters were derived using the Stellar Parameter Classification \citep[SPC;][]{2012Natur.486..375B, 2014Natur.509..593B} tool. TOI-1199 was observed on 7 January 2020 and 9 January 2020 (with 36 and 35 S/N, respectively) and TOI-1273 was observed on 30 December 2019 and 24 January 2020 (with 34 and 37 S/N, respectively). This analysis led to the following parameters: $T_{\mathrm{eff}}=5720\pm50$\,K, $\log g=4.25\pm0.10$, $[Fe/H]=0.50\pm0.08$\,dex and $v\,\sin{i}=3.4\pm0.5$\,\kms for TOI-1199 and $T_{\mathrm{eff}}=5690\pm50$\,K, $\log g=4.35\pm0.10$, $[Fe/H]=0.10\pm0.08$\,dex and $v\,\sin{i}=2.9\pm0.5$\,\kms for TOI-1273. All the values agree within 1$\sigma$ with those reported above from the analysis of SOPHIE spectra. For this four parameters we adopt the weighted arithmetic mean from the two analysis (for $\log g$ we combined the three determinations), which are the values reported in Table~\ref{tab:stellarparam}. 

\begin{figure}
\centering
\includegraphics[width=\hsize]{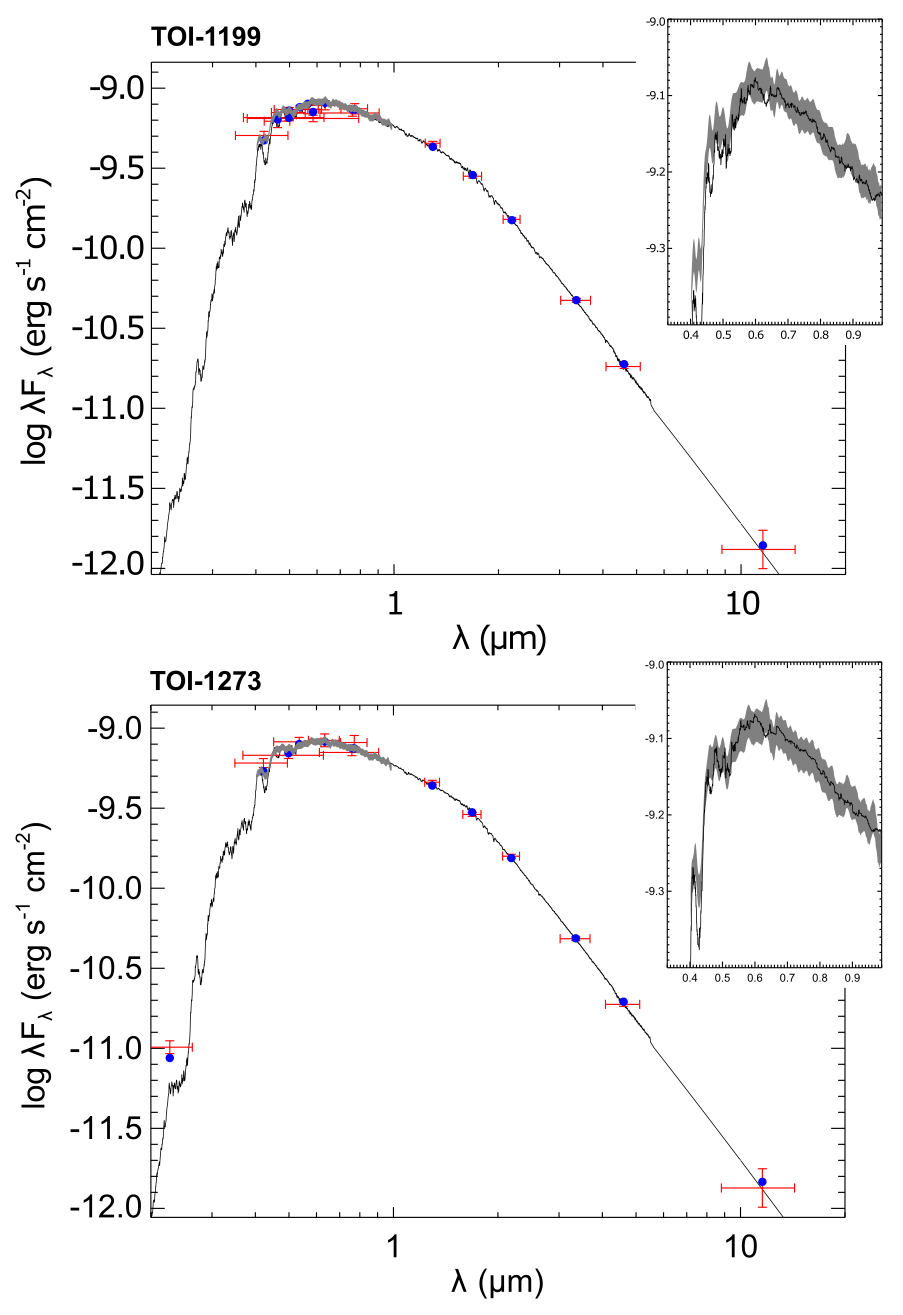}
\caption{SEDs of TOI-1199 and TOI-1273. Red symbols represent the observed photometric measurements; the horizontal bars mark the effective width of the passband. Blue symbols are the model fluxes from the best-fit PHOENIX atmosphere model (black). The absolute flux-calibrated {\it Gaia} spectrum is overlaid on the model SED as a gray swathe in the inset plots.}
\label{fig:sed}
\end{figure}

We performed an analysis of the broadband spectral energy distribution (SED) of the stars together with the \textit{Gaia} DR3 parallax \citep[with no systematic offset applied; see, e.g.,][]{2021ApJ...907L..33S} to determine an empirical measurement of the stellar radius, following the procedures described in \citet{2016AJ....152..180S}, \citet{2017AJ....153..136S}, and \citet{2018ApJ...862...61S}. We pulled the \textit{JHK}$_S$ magnitudes from {\it 2MASS}, the \textit{W1}--\textit{W3} magnitudes from {\it WISE}, the $G_{\rm BP}~G_{\rm RP}$ magnitudes from {\it Gaia}, the $BVg'r'i'$ magnitudes from APASS, and where available the near-UV magnitude from GALEX. Together, the available photometry spans the full stellar SED over at least the wavelength range 0.4--10~$\mu$m and up to 0.2--20~$\mu$m (see Fig.~\ref{fig:sed}).

We then performed a fit using PHOENIX stellar atmosphere models \citep{2013A&A...553A...6H}, with the main parameters being the effective temperature ($T_{\rm eff}$) and metallicity ([Fe/H]), which we adopted from the spectroscopic values, as well as the extinction $A_V$, which we limited to maximum line-of-sight value from the Galactic dust maps of \citet{1998ApJ...500..525S}. The resulting fits (Fig.~\ref{fig:sed}) have a best-fit $A_V = 0.02 \pm 0.02$ and $0.01 \pm 0.01$ for TOI-1199 and TOI-1273, respectively, with a reduced $\chi^2$ of 1.2 and 1.4, respectively. Integrating the (un-reddened) SED gives the bolometric flux at Earth, $F_{\rm bol} = 1.0553 \pm 0.0017 \times 10^{-9}$ erg~s$^{-1}$~cm$^{-2}$ and $1.097 \pm 0.013 \times 10^{-9}$ erg~s$^{-1}$~cm$^{-2}$, respectively. Taking the $F_{\rm bol}$ together with the {\it Gaia\/} parallax gives the bolometric luminosity directly, $L_{\rm bol} = 2.015 \pm 0.008$~L$_{\rm bol}$ and $1.07 \pm 0.01$~L$_{\rm bol}$, respectively. The stellar radius then follows from the Stefan-Boltzmann relation, giving $R_\star = 1.45 \pm 0.03$~R$_\odot$ and $1.06 \pm 0.02$~R$_\odot$, respectively. In addition, we estimate the stellar mass from the empirical relations of \citet{2010A&ARv..18...67T}, giving $M_\star = 1.23 \pm 0.07$~M$_\odot$ and $1.06 \pm 0.06$~M$_\odot$, respectively.

\begin{figure}
\centering
\includegraphics[width=\hsize]{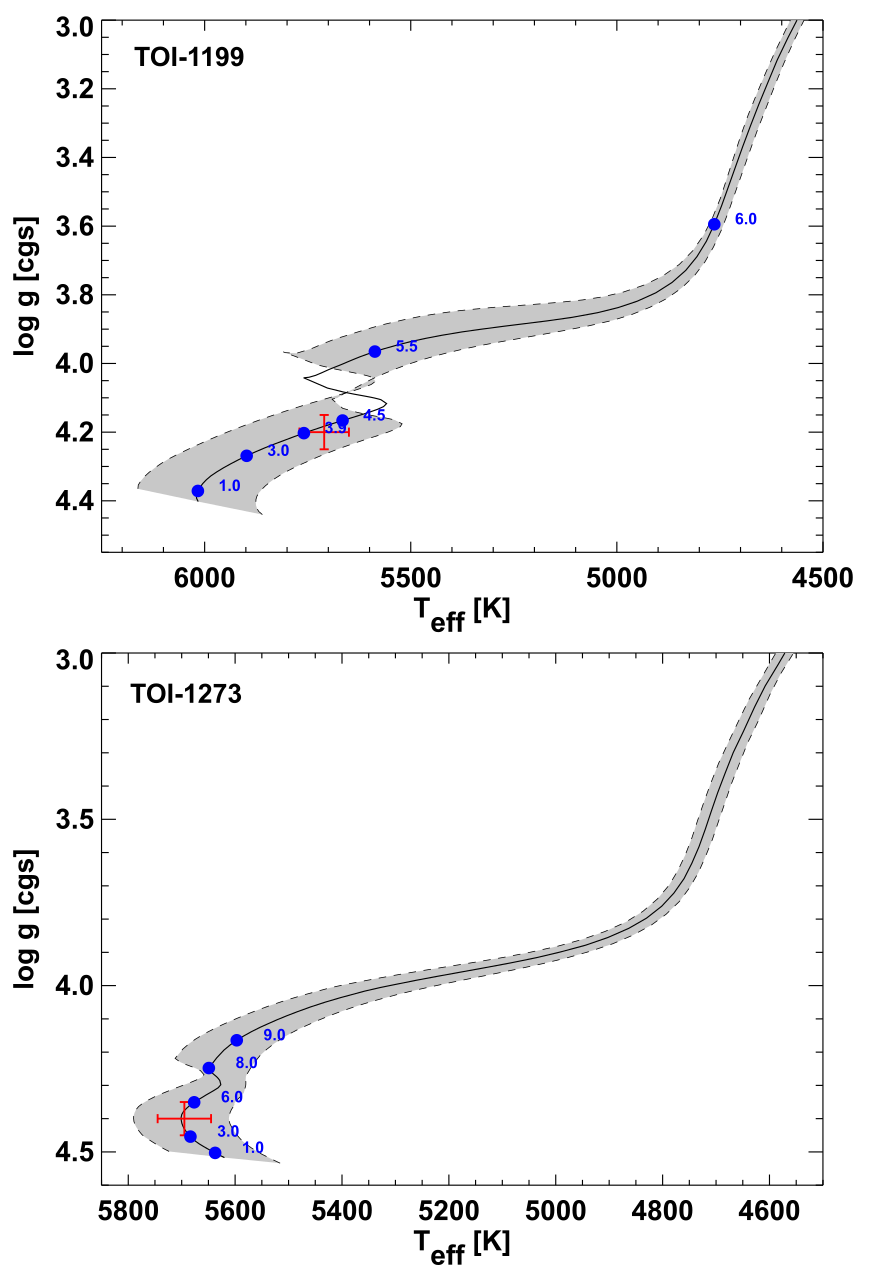}
\caption{Kiel diagrams for TOI-1199 and TOI-1273, showing $\log g$ versus $T_{\rm eff}$ (red symbol) against the Yonsei-Yale stellar evolutionary model for the inferred mass and metallicity (gray swathe). Model ages are represented at various points as blue symbols with labels in Gyr.}
\label{fig:isochrones}
\end{figure}

We also estimate the stellar ages from the empirical rotation-activity-age relations of \citet{2008ApJ...687.1264M}, as follows. We estimate the projected stellar rotation period from the spectroscopic $v\sin i$ together with $R_\star$, giving $P_{\rm rot}/\sin i = 21.9 \pm 0.6$~d and $20.7 \pm 2.4$~d, respectively, which imply gyrochronological ages of $4.2 \pm 0.2$~Gyr and $2.1 \pm 0.5$~Gyr, respectively. At the same time, the spectroscopically determined chromospheric activity indices, $R'_{\rm HK}$, imply ages of $8.4 \pm 3.6$~Gyr and $6.6 \pm 3.2$~Gyr, respectively, and predict rotation periods of $23.1 \pm 5.4$~d and $42 \pm 11$~d, respectively. However, given the poor constraint on log $R'_{\rm HK}$ this estimation is the least reliable. Finally, placing the stars in a Kiel diagram (Fig.~\ref{fig:isochrones}) against the Yonsei-Yale stellar evolutionary models implies ages of $4.2 \pm 0.3$~Gyr and $4.5 \pm 1.5$~Gyr, respectively. The age reported on Table~\ref{tab:stellarparam} in the case of TOI-1199 is the weighted arithmetic mean between the gyrochronological and the Kiel diagram ages, resulting in an estimated age of $4.2 \pm 0.2$~Gyr. For TOI-1273 we did the same but we also added a systematic error to each measurement considering that the difference in the values from both methods imply an underestimation of the error bars, the final estimated age is $3.1 \pm 1.6$~Gyr.

Something to remark is that both stars are solar analogs, the main difference being the really high metallicity of TOI-1199. TOI-1273 has parameters consistent within 2 and 3$\sigma$ to those of the Sun, making it a candidate to solar twin. This makes these two targets more interesting since solar analogs and especially solar twins offer an opportunity to derive more accurate stellar parameters and consequently more accurate planetary parameters. The adopted stellar parameters for both stars are summarized in Table~\ref{tab:stellarparam}. 

\subsection{Planetary parameters}
\label{planetarysubsection}
    \begin{figure*}
    \centering
    \resizebox{\hsize}{!}
            {\includegraphics{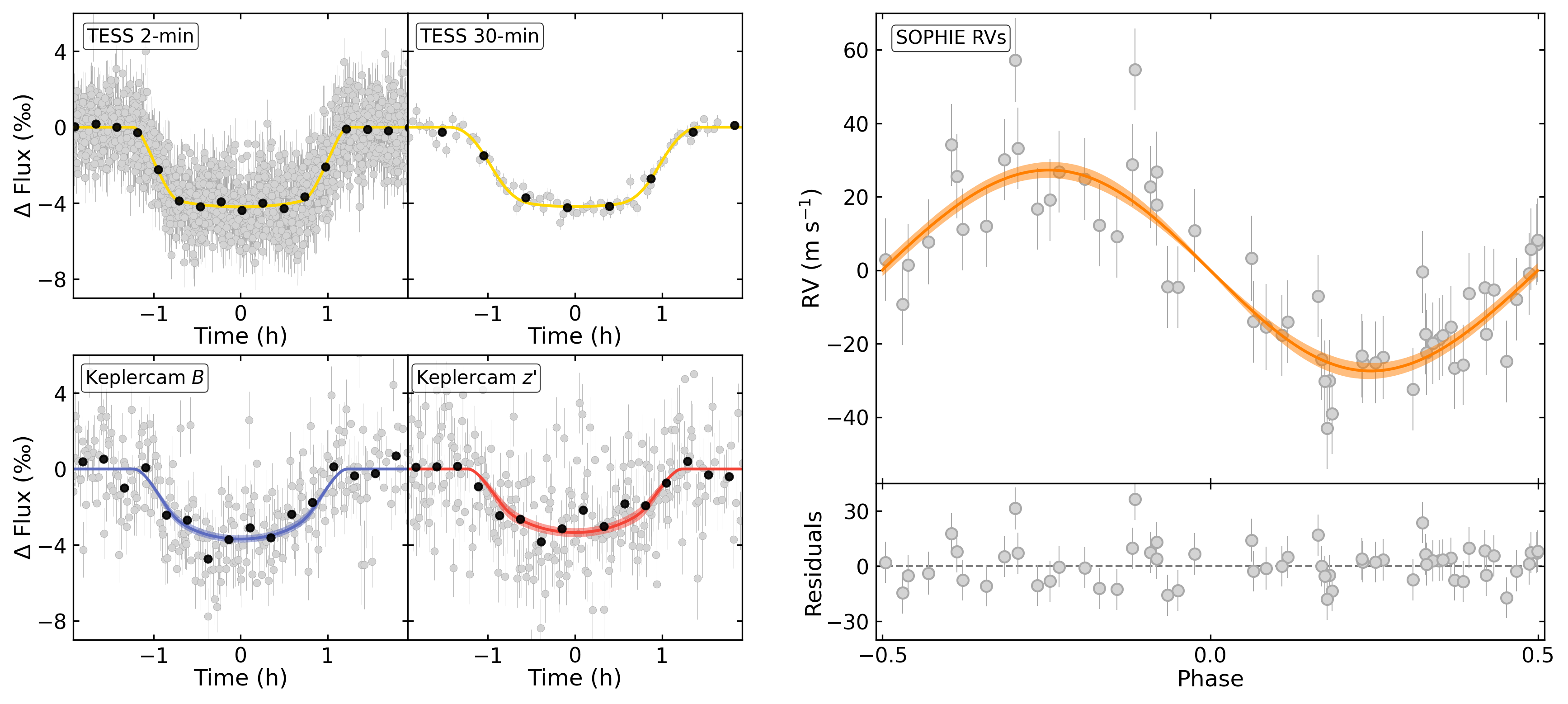}}
            \caption{Best-fit model and residuals for TOI-1199. On each curve, the solid colored line and its overlay correspond to the median and 16th-84th percentile regions of the solution, respectively. \textit{Left:} Folded light curves used in the model. White markers show the binned light curve with a bin size of 0.01\,d. \textit{Right panel:} Phase-folded RVs and residuals.}
    \label{fig:fit1199}
    \end{figure*}

    \begin{figure*}
    \centering
    \resizebox{\hsize}{!}
            {\includegraphics{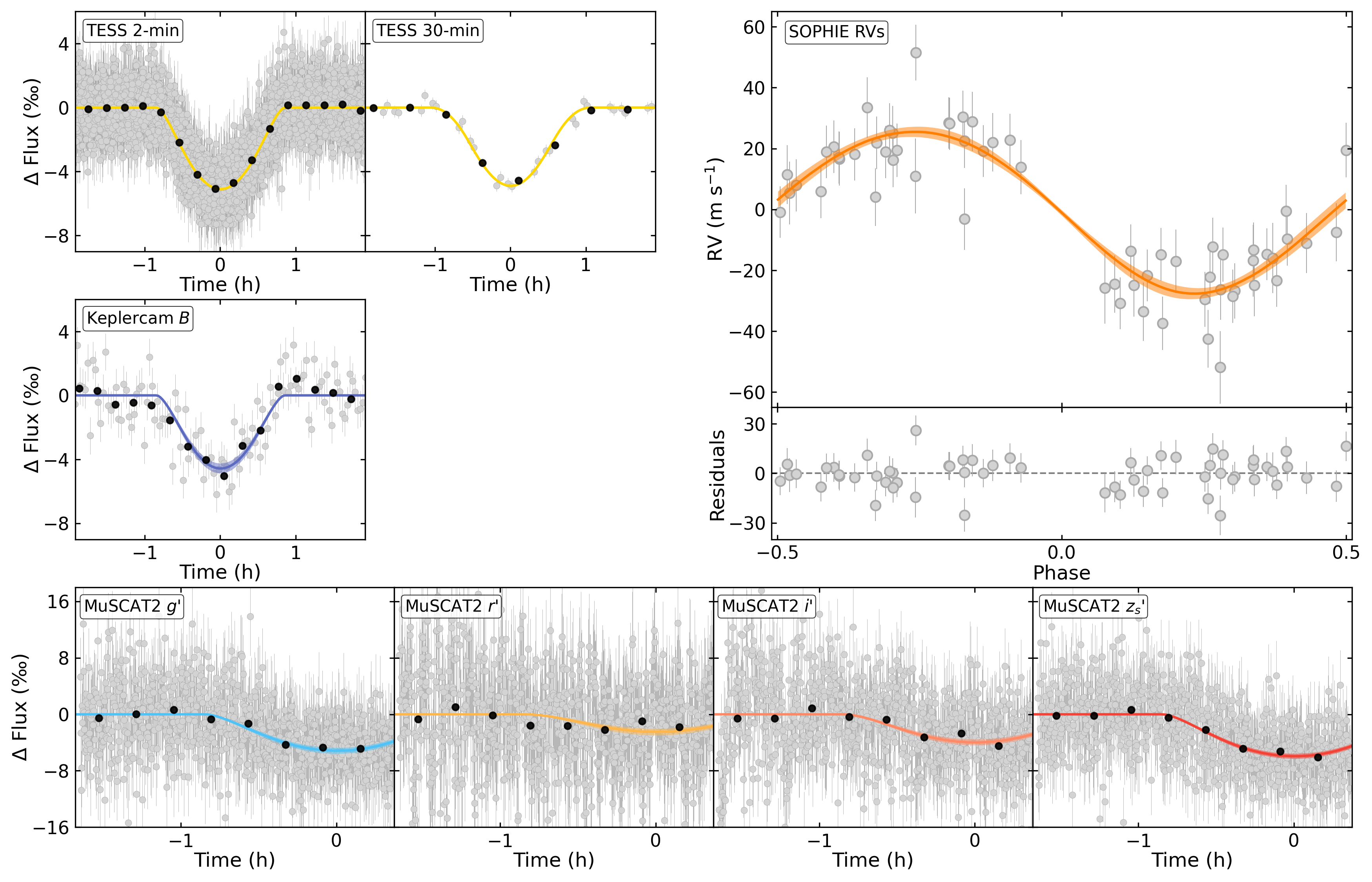}}
            \caption{Best-fit model and residuals for TOI-1273.  On each curve, the solid colored line and its overlay correspond to the median and 16th-84th percentile regions of the solution, respectively. \textit{Left and bottom panel:} Folded light curves used in the model. White markers show the binned light curve with a bin size of 0.01\,d. \textit{Right panel:} Phase-folded RVs and residuals.}
    \label{fig:fit1273}
    \end{figure*}
    
We fit a joint model to the photometric and RV observations of both targets. For TOI-1199 the inputs of the joint model are the 60 SOPHIE RVs, the light curves from four TESS sectors (two 30-minute cadence and two 2-minute cadence) and the KeplerCam light curves in bands \textit{B} and \textit{z'} (Fig.~\ref{fig:fit1199}). In the case of TOI-1273, we used 60 SOPHIE RVs, TESS light curves from six sectors (five high cadence and one low cadence), one KeplerCam \textit{B}-band light curve, and four MuSCAT2 light curves in bands \textit{g'}, \textit{r'}, \textit{i',} and \textit{z'} (Fig.~\ref{fig:fit1273}). In the two models we only used the points within 0.1\,d from the mid transits. 

Bayesian inference was carried out to obtain the planetary parameters from a probabilistic model, and it was similar for both targets. We used the \texttt{EXOPLANET} Python package \citep{2021JOSS....6.3285F}, which is built within the framework of \texttt{PyMC3}, a flexible and open source Python probabilistic programming language with a built-in Hamiltonian Monte Carlo Sampler \citep{2016ascl.soft10016S}. One advantage of using the \texttt{EXOPLANET} toolkit is that it has many built-in common functions and tools for modeling exoplanetary time series like the solver of the Kepler equation and others.

The prior distributions defined for the parameters of the joint models are shown in Appendix~\ref{apx:priors}. The stellar mass and radius have normal priors informed by the analysis presented in Sect.~\ref{subsec:stellarparam} and bounded between $0$ and $3$. For the limb darkening coefficients $(q_1$ and$ \,q_2$), we used for each bandpass the two-parameter quadratic law parametrization described by \citet{2013MNRAS.435.2152K}. Additionally, a mean baseline flux and a jitter term is defined for each instrument and bandpass (in the case of TESS also for each cadence), the jitter is introduced in the model by adding it in quadrature to the error of the light curves. From inspecting the transits on Fig.~\ref{fig:fit1273}, high values for the impact parameters are expected, especially for TOI-1273 where a clear V-shape is observed indicating a grazing transit. Considering this, we set a uniform prior between $0$ and 1\,$+$\,$R_p/R_\star$ to the impact parameter $b$ to make sure we include the case of $b$\,>\,1 in the parameter space. For the ratio of radius $R_p/R_\star$, we defined a wide log-normal prior that was informed by the measured depth of the transits. In the case of TOI-1273 we also included a lower bound constraint to the planetary bulk density because of the grazing transit, this is discussed in detail in Sect.~\ref{subsec:prior1273}.

To account for any possible long-term trends in the RV curves we included a second degree polynomial of time with the three coefficients as free parameters, which are referred as {RV trend 2}, {RV trend 1}, and {RV trend 0} in Table~\ref{tab:priors2} and are the quadratic, linear, and constant coefficients, respectively. Additionally, an RV jitter term is added with a log-normal prior, and for the RV semi-amplitude $K$ we set log-normal priors informed by the estimated value from a preliminary fit on the RV curves. The time of mid transit, $T_0,$ and the orbital period, $P,$ have priors informed by the values obtained from the box least-squares analysis. 
Finally, because it is well known that sampling directly the eccentricity and the argument of periastron can be problematic for most Markov chain Monte Carlo (MCMC) samplers \citep{2018haex.bookE.149P}, we sampled for $\sqrt{e} \sin{\omega}$ and $\sqrt{e} \cos{\omega}$ instead with a uniform prior within a unit disk, which leads to a uniform prior on $e$ as noted by \citet{2011ApJ...726L..19A}. 

After defining the models, we calculated the local maximum a posteriori solution, which was then used as a starting point for the No-U-Turn Sampler, a variation of a Hamiltonian Monte Carlo method described by \citet{2011arXiv1111.4246H}. We sampled the posterior distributions of the parameters with 4000 tuning steps, 4000 draws and 2 independent chains for each model. The MCMCs do not show convergence problems, the Gelman-Rubin statistic is close to 1 for all parameters and well below $1.1$, which is generally considered the threshold to indicate convergence problems \citep{gelman2013bayesian}.

For TOI-1199\:b we obtain a well-constrained radius of $R_P$\,=\,0.938\,$\pm$\,0.025\,\RJ\, and a mass of $M_P$\,=\,0.239\,$\pm$\,0.020\,\MJ, resulting in a bulk density of 0.358\,$\pm$\,0.041\,\gcc. For TOI-1273\:b we constrain its radius to $R_P$\,=\,0.99\,$\pm$\,0.22\,\RJ, where the larger uncertainty is a consequence of the grazing transit. Its mass is well determined to $M_P$\,=\,0.222\,$\pm$\,0.015\,\MJ, giving a bulk density of  0.28$\,\pm\,$0.11\,\gcc. Equilibrium temperatures ($T_\mathrm{eq}$) are calculated for both asuming zero albedo and full day-night heat distribution according to
\begin{equation} \label{eq:teq}
        T_\mathrm{eq} = T_\star \sqrt{\frac{R_\star}{a}} \left(\frac{1}{4}\right)^{1/4} \\
.\end{equation}
With these masses and short periods, both planets are located in the hot Saturn parameter space. 

TOI-1199\:b has an eccentricity of 0.030$\,\pm\,$0.029, which is compatible with zero within 2$\sigma$ and with a 3$\sigma$ upper limit of $0.14$. For TOI-1273\:b we found a slightly larger value of 0.055$\,\pm\,$0.032, also compatible with zero at 2$\sigma$ and with a 3$\sigma$ upper limit of $0.15$. Considering the low values of eccentricity obtained for both planets we did a run for each fixing $e=0$ to simplify the model and see if this might improve the constraints on other parameters. The results agreed within 1$\sigma$ for all parameters in both targets, we therefore decided to let the eccentricity as a free parameter.

With the available data we find no evidence of additional companions. As shown on Fig.~\ref{fig:periodograms}, the residuals of the models do not present any further periodic signals. This allows us to discard other short-period planets at least with semi-amplitudes larger than the combined uncertainty from the RV measurements and the RV jitter; this value is 11\,\ms\, for TOI-1199 and 9\,\ms\, for TOI-1273. Furthermore, the second degree polynomial model provides upper limits for any possible long-term trend. If we assume a circular orbit for a potential outer companion, we can discard planets with semi-amplitudes larger than the RV variation present in the polynomial model, and periods shorter than two times the time-span. With this criteria, we can exclude for TOI-1199 planets with periods shorter than 1800 days and semi-amplitudes larger than 33\,\ms\,with 3$\sigma$ confidence, which translates to a lower limit mass of 2.3\,\MJ. And for TOI-1273 the excluded companions have periods shorter than 1800 days and semi-amplitudes larger than 23\,\ms\,at 3$\sigma$, corresponding to a lower limit mass of 1.6\,\MJ. Additionally, we checked the \textit{Gaia} DR3 renormalized unit weight error (RUWE) and astrometric excess noise (AEN) for TOI-1199 ($\rm{AEN} = 0.110$ mas, $\rm{RUWE} = 1.12$) and TOI-1273 ($\rm{AEN} = 0.056$ mas, $\rm{RUWE} = 0.86$). Neither show evidence of an accelerated motion signal beyond a 95\% confidence level. Thus, there are no hints of a stellar or brown dwarf companion on either object. Maintaining a long-term RV monitoring of both stars is recommended to keep searching for possible companions and we are currently continuing such observations with SOPHIE.

Table~\ref{tab:planets} shows the orbital and physical parameters obtained for the two planets. The full list of parameters from the joint models is shown in Appendix~\ref{apx:priors} and the corner plots of the MCMC sampling of the posterior distributions are shown in Appendix~\ref{apx:cornerplots}.

\begin{table}
\caption{Fitted and derived planetary parameters.}             
\centering          
\begin{tabular}{l c c}     
\hline\hline        
Parameter & TOI-1199\:b & TOI-1273\:b \\
\hline
\multicolumn{3}{l}{\textit{Fitted parameters}} \\
    $P$\,(d) & $3.671463\pm0.000003$ & $4.631296\pm0.000003$ \\
    $T_0$\,(TBJD) & $2420.5376\pm0.0004$ & $1712.3468\pm0.0004$ \\
    $b$ & $0.849\pm0.016$ & $0.958\pm0.030$ \\
    $K$\,(\ms) & $27.5\pm2.1$ & $26.7\pm1.5$ \\
    $\sqrt{e} \sin{\omega}$  & $-0.01\pm0.12$ & $-0.18\pm0.11$ \\
    $\sqrt{e} \cos{\omega}$  & $0.06\pm0.14$ & $0.04\pm0.14$ \\
    $R_{p}/R_{\star}$ & $0.0663\pm0.0012$ & $0.096\pm0.021$ \\
\hline
\multicolumn{3}{l}{\textit{Derived parameters}} \\
    $M$\,(\MJ) & $0.239\pm0.020$ & $0.222\pm0.015$ \\
    $R$\,(\RJ) & $0.938\pm0.025$ & $0.99\pm0.22$ \\
    $e$ & $0.030\pm0.029$ & $0.055\pm0.032$ \\
    $a$\,(AU) & $0.04988\pm0.00091$ & $0.0549\pm0.0010$ \\
    $\rho~$ (\gcc) & $0.358\pm0.041$ & $0.28\pm0.11$ \\
    $T_{\text{eq}}~$(K) & $1486\pm20$ & $1211\pm15$ \\
    $\omega$ (rad) & $1.0\pm1.7$ & $1.9\pm2.4$ \\
\hline                  
\end{tabular}
\label{tab:planets}
\end{table}

\subsection{Prior TOI-1273\:b density constraint  }
\label{subsec:prior1273}
Some problems arise with grazing orbits. For example, it causes otherwise uncorrelated parameters $b$ and $R_P/R_\star$ to become highly correlated, making the sampling of this parameters significantly more difficult and also producing higher uncertainties on the derived planet radius and impact parameter. This problem cannot be overcome with re-parametrization. Another thing to consider is that since the transit is so close to the edge of the stellar disk, the model becomes more sensitive to the limb darkening parameters, and as established by \citet{2013A&A...560A.112M} the theoretical knowledge of the limb darkening close to the edge is inaccurate.

\begin{figure}
\centering
\includegraphics[width=\hsize]{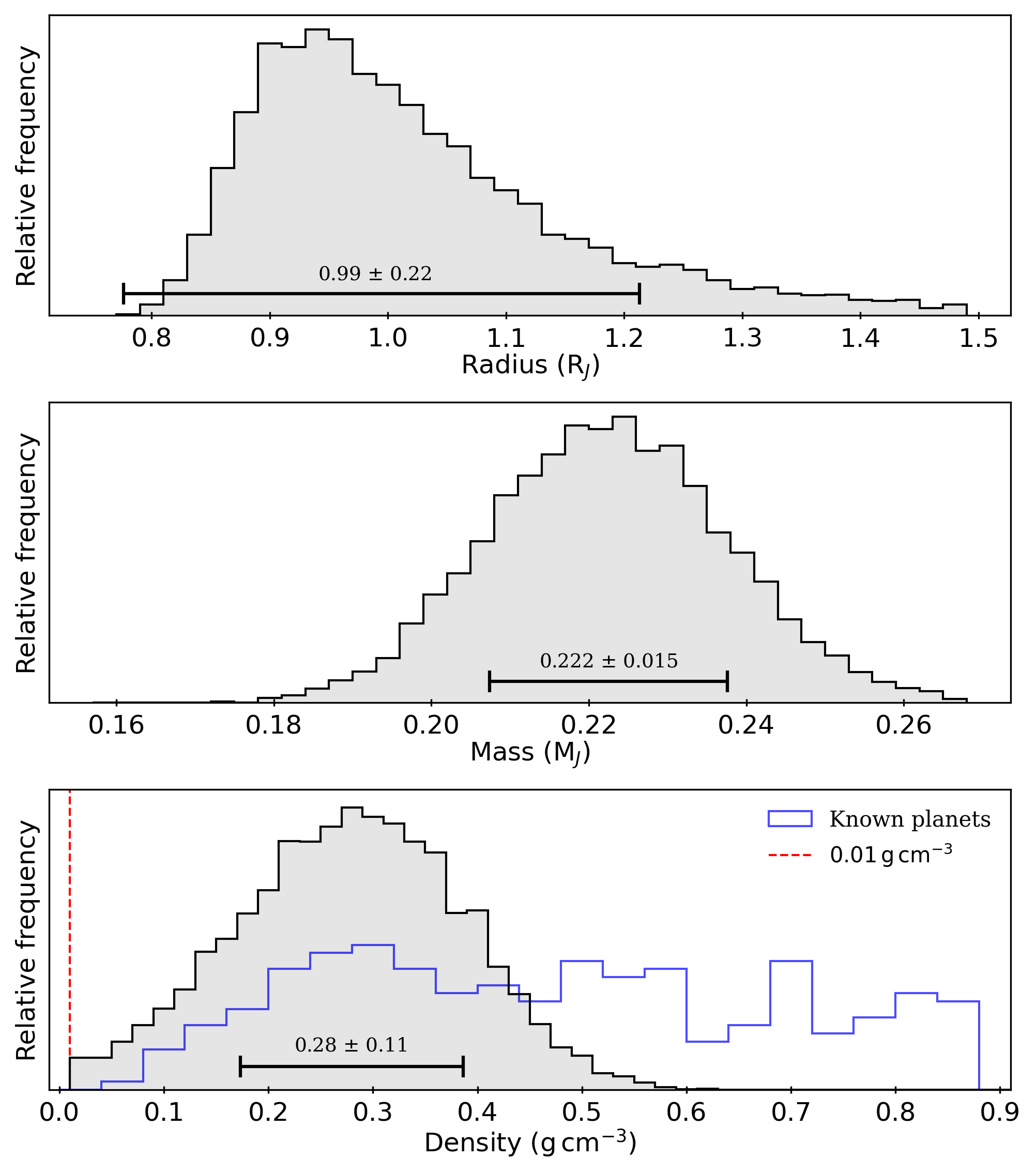}
\caption{Histograms from the posterior distribution samples of TOI-1273\:b (radius, mass, and density from top to bottom). The error bars represent the reported value for each parameter, which is the median $\pm$\,1$\sigma$. The red dashed line on the lower panel shows the value at which we put the prior constraint on the density. In the lower panel the bulk density distribution of confirmed planets with masses and radii determined to better than 20\% precision is also shown.}
\label{fig:densityhist}
\end{figure}

Traditional random-walk based MCMC algorithms become inefficient with correlated parameters. Although this is less of a problem for the NUTS sampler \citep{2016ascl.soft10016S}, highly correlated parameters remain problematic. In an initial run of the model two problems became evident for TOI-1273\:b due to the degeneracy between $b$ and $R_P/R_\star$: (1) the effective sample size for this two parameters was significantly smaller than on TOI-1199\:b; and (2) the two parameters were not well constrained in the parameter space, which resulted in having many samples located in physically unrealistic regions (that is, transits of impossibly big planets for the given mass with consequently large impact parameters, and extremely low densities). To overcome this we set a  prior constraint on the planetary density. We limited the model to systems with bulk densities greater or equal to 0.01\,\gcc, we consider this value a reasonable limit for the density. Figure \ref{fig:densityhist} shows the posterior distributions for the radius, mass and density of TOI-1273\:b, where the cut at 0.01\,\gcc~is visible in the lower panel. The distribution of bulk densities for all known exoplanets with masses and radii known to better than 20\% precision is also shown, where we see that there are no planets with bulk densities lower than 0.01\,\gcc. When applying this restriction to the model we get a better constraint to the radius of TOI-1273\:b. The results reported in Tables~\ref{tab:planets}, \ref{tab:priors}, and \ref{tab:priors2} are from the model with this restriction.

\section{Discussion}
\label{sec:discussion}
\subsection{Radius--mass diagram}
\begin{figure*}
\centering
\includegraphics[width=\hsize]{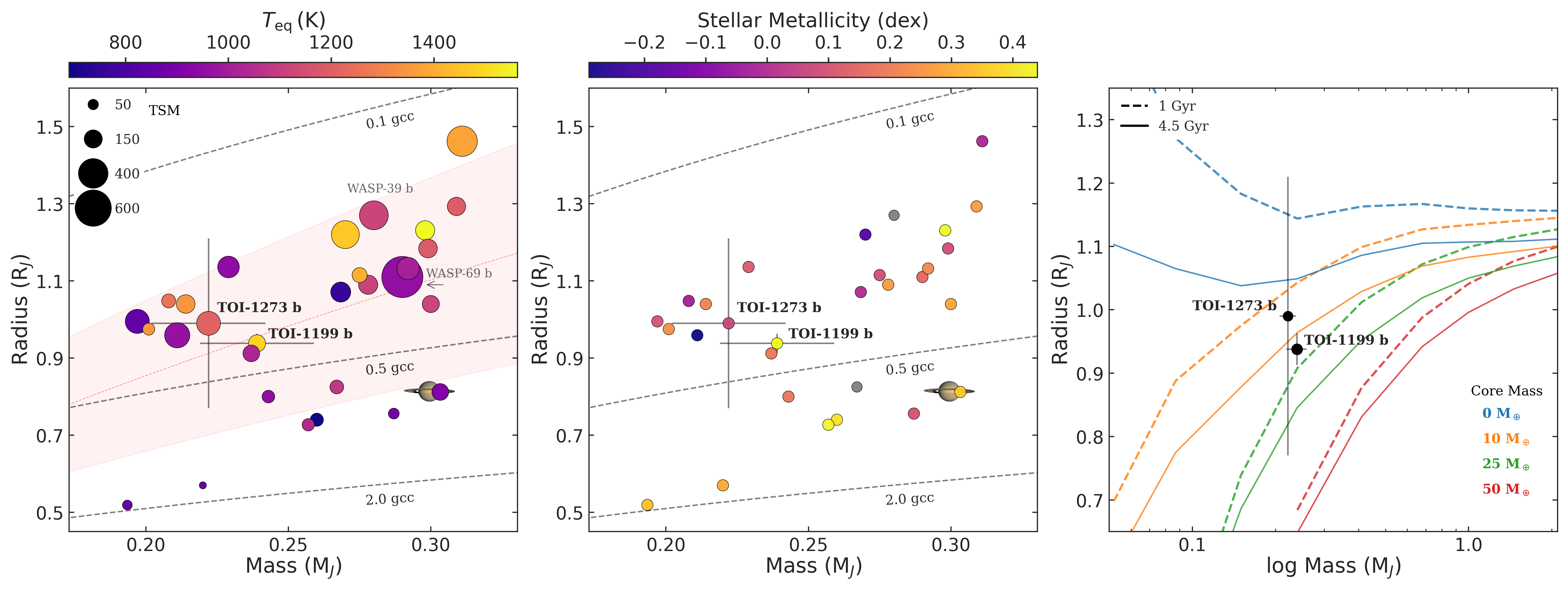}
\caption{Radius-mass diagrams. \textit{Left panel:} Planetary mass and radius of the confirmed planet population with 1\,<\,$P$\,<\,10\,d and masses 60\,<\,$M_P$\,<\,100\,\ME, with radii and masses determined to better than 20\% precision. The gray dashed lines show constant density values of 0.1, 0.5 and 2.0\,\gcc. The position of Saturn is shown for comparison. The equilibrium temperatures for each planet are color coded and the markers are sized according to their TSM \citep{2018PASP..130k4401K}. The R-M empirical relation found by \citet{2020A&A...634A..43O} is shown by a dashed red  line with the $\pm$1$\sigma$ regions colored. Two similar planets for which atmospheric characterization has been achieved are shown, namely WASP-39\,b \citep{2023Natur.614..659R} and WASP-69\,b \citep{2021A&A...656A.142K, 2023A&A...673A.140L, 2023MNRAS.521.5860O}. \textit{Middle panel:} Same sample shown color coded by the metallicity of their host stars. Planets for which no host star metallicity determination is available are shown in gray. \textit{Right panel:} Masses and radii of TOI-1199\,b and TOI-1273\,b shown along the structural models of \citet{2007ApJ...659.1661F} for planets at 0.045 AU.}
\label{fig:massradius}
\end{figure*}

The combined analysis of high-resolution spectroscopy and space- and ground-based photometry allowed the determination of the masses and radii of planetary companions for the two target stars. Their masses and radii are $M_P$\,=\,0.239\,$\pm$\,0.020\,\MJ\   and $R_P$\,=\,0.938\,$\pm$\,0.025\,\RJ\ for TOI-1199\,b; and $M_P$\,=\,0.222\,$\pm$\,0.015\,\MJ\ and  $R_P$\,=\,0.99\,$\pm$\,0.22\,\RJ\ for TOI-1273\,b. The physical properties of both planets are similar to those of Saturn ($M_{\textrm{Saturn}}$\,=\,0.30\,\MJ, $R_{\textrm{Saturn}}$\,=\,0.83\,\RJ, and $\rho_{\textrm{Saturn}}$\,=\,0.68\,\gcc) but they lie on short-period orbits (3.67\,d and 4.63\,d, respectively) and are less dense (0.358$\,\pm\,$0.041\,\gcc\,and 0.28$\,\pm\,$0.11\,\gcc, respectively). To put them in context, we compared their masses and radii to the known exoplanet population in the range 1\,<\,$P$\,<\,10\,d and 60\,<\,$M_P$\,<\,100\,\ME\, with masses and radii determined to better than  20\% precision (left and middle panels of Fig.~\ref{fig:massradius}). We can see that both planets are among the group with the lower densities for this range of masses and periods. In the left panel, the empirical radius-mass relation for volatile rich planets ($\rho$\,<\,3.3\,\gcc) found by \citet{2020A&A...634A..43O} is also shown, where we note that the masses and radii of TOI-1199\,b and TOI-1273\,b are well described within 1$\sigma$ by this relation. In the middle panel of Fig.~\ref{fig:massradius}, TOI-1199\,b stands out as one of the planets with the highest host star metallicity among similar known systems. For planets in this range of masses, \citet{2012A&A...540A..99E} finds a negative correlation between host star metallicity and radius, this tendency is more or less visible here.
We also compared the mass and radius of our planets to the mass-radius predictions for giant irradiated planets by \citet{2007ApJ...659.1661F}, the right panel of Fig.~\ref{fig:massradius} shows the theoretical models with different core masses and ages at a fixed semimajor axis of 0.045\,AU ($a = 0.0492 \pm 0.0003$ for TOI-1199 b and $a = 0.0540 \pm 0.0004$ for TOI-1273 b). These models assume a heavy elements solid core and a H/He envelope. While the large uncertainty on the radius of TOI-1273\,b makes poor the comparison with the models, we see that for the radius, mass and age ($4.2 \pm 0.2$ Gyr) of TOI-1199\,b, the planet is best described by a structure with a core mass between 10 and 25 \ME.
\subsection{Position in the Neptunian desert}
\begin{figure}
\centering
\includegraphics[width=\hsize]{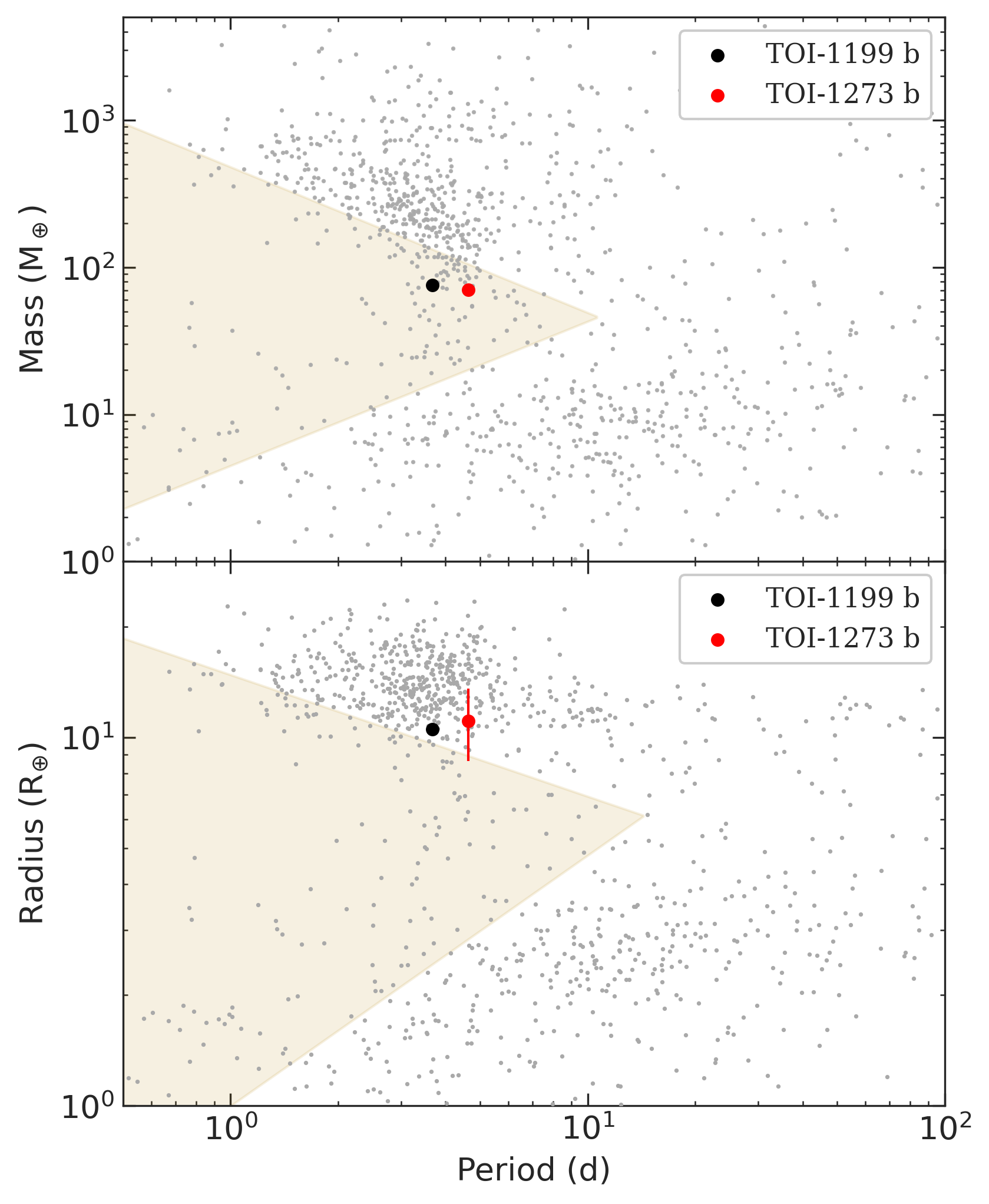}
\caption{TOI-1199\,b (\textit{black marker}) and TOI-1273\:b (\textit{red marker}) in the context of the Neptunian desert regions in period versus mass (\textit{top panel}) and period versus radius (\textit{bottom panel}) as defined by \citet{2016A&A...589A..75M}.}
\label{fig:desert}
\end{figure}

The lack of Neptune-sized planets in closed-in orbits is known as Neptunian desert, whose boundaries were determined by \citet{2016A&A...589A..75M}. Since the principal methods of exoplanet detection should be highly efficient to detect this kind of object, explaining the origin of the desert is an intriguing subject of investigation in the field. It represents an opportunity to explore the possible underlying physical processes that cause it. In Fig.~\ref{fig:desert} we show the position of TOI-1199\,b and TOI-1273\,b within the mass-period (top) and radius-period (bottom) boundaries. \citet{2018MNRAS.479.5012O} argue that the lower boundary can be explained by photoevaporation, while the upper boundary is supposed to be caused by high eccentricity migration followed by tidal disruption. But \citet{2023ApJ...945L..36T} arrive at a slightly different conclusion, where photoevaporation can account at least partially on carving out the upper boundary. Mass loss rates have been surveyed on gas giants at the edges of the desert to probe the different paths on planet formation and evolution that lead to the observed population. For instance, \citet{2022AJ....164..234V} find stability against photoevaporation on a sample of seven planets (with six in the upper edge). They conclude that other mechanisms must be responsibly for the upper edge of the desert. While this discussion is out of the scope of this work, we provide two new hot Saturns on the upper edge with masses determined to better than 10\% precision, which will be useful for future studies on the functional dependence of the upper boundary on stellar and planetary properties.

\subsection{Transmission spectroscopy metric}
The low densities of these planets indicate the presence of gaseous envelopes, which combined with the bright target stars ($V$\,$\approx$11\,mag) makes the prospect of atmospheric characterization worth examining. We calculated their transmission spectroscopy metric\footnote{The TSM is defined in the sub-Jovian population for radius between 4 and 10\,\RE. We extrapolated this to the slightly larger than 10 \RE \ radii of our planets.} \citep[TSM;][]{2018PASP..130k4401K} obtaining a value of 134\,$\pm$\,17 for TOI-1199\,b and 261\,$\pm$\,175 for TOI-1273\,b. The large uncertainty for TOI-1273\:b comes from  TSM\,$\propto$\,$R_{\mathrm{P}}^3$, and for this target the metric is overestimated  given that as a cause of the grazing transit not all its atmosphere transits the star. However, the value for TOI-1199\,b is well determined and is above the threshold of 90 set by \citet{2018PASP..130k4401K} for sub-Jovians to be considered high quality atmospheric characterization targets. 
\section{Conclusions}
\label{sec:conclusion}

In this work we have reported the discovery and characterization of two new transiting hot Saturns orbiting the nearby and bright G-type stars TOI-1199 and TOI-1273, located at 247.0$^{+0.8}_{-0.7}$\,pc and 176.0$^{+0.4}_{-0.4}$\,pc from the Sun, respectively. The joint analysis of TESS and ground-based photometry and RVs from SOPHIE spectroscopic follow-up allowed the validation of the planets and the determination of their orbital and physical properties. TOI-1199\:b orbits its host star in a circular compatible orbit ($e$\,$=$\,0.030\,$\pm$\,0.029) with a period of 3.67\,d and has a mass of 0.239\,$\pm$\,0.020\,\MJ\, and a radius of 0.938\,$\pm$\,0.025\,\RJ, giving it a bulk density of $\rho$\,=\,0.358\,$\pm$\,0.041\,\gcc. TOI-1273\:b is also in a circular compatible orbit ($e$\,=\,0.055\,$\pm$\,0.032) with a period of 4.63\,d; it has a mass of 0.222\,$\pm$\,0.015\,\MJ\, and a radius of 0.99\,$\pm$\,0.22\,\RJ, for a bulk density of $\rho$\,=\,0.28\,$\pm$\,0.11\,\gcc. The bulk densities of the two planets are among the lowest known for their kind. This and the fact that they orbit bright stars and have deep transits makes them interesting candidates for future atmospheric studies. TOI-1273\,b now has a high uncertainty in its radius, which translates to poor constraints on its inferred bulk density, composition, and TSM. TOI-1199\:b, on the other hand, is a solid candidate, with a TSM of 134\,$\pm$\,17 and in orbit around a high-metallicity host star.  

\begin{acknowledgements}
We thank the Observatoire de Haute-Provence (CNRS) staff for their support. This work was supported by the “Programme national de plan\'etologie” (PNP) of CNRS/INSU and CNES.
\\
This article is based on observations made with the MuSCAT2 instrument, developed by ABC, at Telescopio Carlos S\'{a}nchez operated on the island of Tenerife by the IAC in the Spanish Observatorio del Teide. This work is partly financed by the Spanish Ministry of Economics and Competitiveness through grants PGC2018-098153-B-C31.
\\
This work is partly supported by JSPS KAKENHI Grant Numbers JP17H04574, JP18H01265, and JP18H05439, Grant-in-Aid for JSPS Fellows Grant Number JP20J21872, JST PRESTO Grant Number JPMJPR1775, and a University Research Support Grant from the National Astronomical Observatory of Japan (NAOJ).
\\
This research has made use of the Exoplanet Follow-up Observation Program (ExoFOP; DOI: 10.26134/ExoFOP5) website, which is operated by the California Institute of Technology, under contract with the National Aeronautics and Space Administration under the Exoplanet Exploration Program.
\\
Funding for the TESS mission is provided by NASA's Science Mission Directorate. KAC acknowledge support from the TESS mission via subaward s3449 from MIT.
\\
Antonio Frasca acknowledges the support from Italian Ministero dell'Universit\'a e Ricerca through the project PRIN-INAF 2019 “Spectroscopically Tracing the Disk Dispersal Evolution".
\\
This work made use of \texttt{tpfplotter} by J. Lillo-Box (publicly available in www.github.com/jlillo/tpfplotter), which also made use of the python packages \texttt{astropy}, \texttt{lightkurve}, \texttt{matplotlib} and \texttt{numpy}.
\\
Some of the observations in this paper made use of the High-Resolution Imaging instrument ‘Alopeke and were obtained under Gemini LLP Proposal Number: GN/S-2021A-LP-105. ‘Alopeke was funded by the NASA Exoplanet Exploration Program and built at the NASA Ames Research Center by Steve B. Howell, Nic Scott, Elliott P. Horch, and Emmett Quigley. ‘Alopeke was mounted on the Gemini North telescope of the international Gemini Observatory, a program of NSF’s OIR Lab, which is managed by the Association of Universities for Research in Astronomy (AURA) under a cooperative agreement with the National Science Foundation. on behalf of the Gemini partnership: the National Science Foundation (United States), National Research Council (Canada), Agencia Nacional de Investigación y Desarrollo (Chile), Ministerio de Ciencia, Tecnología e Innovación (Argentina), Ministério da Ciência, Tecnologia, Inovações e Comunicações (Brazil), and Korea Astronomy and Space Science Institute (Republic of Korea).
\\
E.M. acknowledges funding from FAPEMIG under project number APQ-02493-22 and research productivity grant number 309829/2022-4 awarded by the CNPq, Brazil.
\\
We acknowledge funding from the French National Research Agency (ANR) under contract number ANR-18-CE31-0019 (SPlaSH).
\\
We acknowledge funding from the French National Research Agency in the framework
of the Investissements d Avenir program (ANR-15-IDEX-02), through the funding of the “Origin of Life” project of the Grenoble-Alpes University.
\\
NCS was acknowledges the funding by the European Union (ERC, FIERCE, 101052347). Views and opinions expressed are however those of the author(s) only and do not necessarily reflect those of the European Union or the European Research Council. Neither the European Union nor the granting authority can be held responsible for them. This work was supported by FCT - Fundação para a Ciência e a Tecnologia through national funds and by FEDER through COMPETE2020 - Programa Operacional Competitividade e Internacionalização by these grants: UIDB/04434/2020; UIDP/04434/2020.
\\
This paper made use of data collected by the TESS mission and are publicly available from the Mikulski Archive for Space Telescopes (MAST) operated by the Space Telescope Science Institute (STScI). Funding for the TESS mission is provided by NASA’s Science Mission Directorate. We acknowledge the use of public TESS data from pipelines at the TESS Science Office and at the TESS Science Processing Operations Center. Resources supporting this work were provided by the NASA High-End Computing (HEC) Program through the NASA Advanced Supercomputing (NAS) Division at Ames Research Center for the production of the SPOC data products.
\\
We acknowledge financial support from the Agencia Estatal de Investigaci\'on of the Ministerio de Ciencia e Innovaci\'on MCIN/AEI/10.13039/501100011033 and the ERDF “A way of making Europe” through project PID2021-125627OB-C32, and from the Centre of Excellence “Severo Ochoa” award to the Instituto de Astrofisica de Canarias.
\\
This work is partly supported by JSPS KAKENHI Grant Number JP18H05439 and JST CREST Grant Number JPMJCR1761.
\\
This article is based on observations made with the MuSCAT2 instrument, developed by ABC, at Telescopio Carlos Sánchez operated on the island of Tenerife by the IAC in the Spanish Observatorio del Teide.
\end{acknowledgements}

\bibliographystyle{aa} 
\bibliography{aa} 

\begin{appendix}

\section{SOPHIE spectroscopy}
\label{apx:rvs}
In this appendix we present the RVs and BISs time series derived from SOPHIE spectroscopy for TOI-1199 and TOI-1273.

\begin{table}[h!]
    \small
    \caption{SOPHIE RV and BIS time series for TOI-1199.}   
    \label{tab:1199_rvs}      
    \centering          
    \begin{tabular}{l l l l }     
    \hline\hline       
    Time (RJD) & RV ($ms^{-1}$) & $\sigma_{RV}$ ($ms^{-1}$) & BIS ($ms^{-1}$)\\ 
    \hline                    
    58856.67683 & -18.8804 & 0.0048 & 0.0276 \\
    58860.5179 & -18.8711 & 0.0045 & 0.0204 \\
    58885.63826 & -18.8963 & 0.0048 & 0.0301 \\
    58887.55416 & -18.8517 & 0.0046 & 0.0187 \\
    58916.45412 & -18.8075 & 0.005 & 0.0046 \\
    59170.69638 & -18.8745 & 0.0044 & 0.0177 \\
    59183.71984 & -18.8631 & 0.0044 & 0.0141 \\
    59205.70776 & -18.8713 & 0.0044 & 0.0355 \\
    59206.69469 & -18.8512 & 0.0045 & 0.0239 \\
    59245.58813 & -18.8895 & 0.0045 & 0.0108 \\
    59247.64243 & -18.8481 & 0.0045 & 0.0408 \\
    59269.48441 & -18.8618 & 0.0045 & -0.012 \\
    59273.59262 & -18.8602 & 0.005 & 0.0741 \\
    59274.53395 & -18.896 & 0.0044 & 0.0057 \\
    59275.47408 & -18.8653 & 0.0045 & 0.0165 \\
    59277.58059 & -18.8678 & 0.0055 & 0.0402 \\
    59278.53871 & -18.8715 & 0.0044 & 0.0073 \\
    59279.57273 & -18.8369 & 0.0044 & -0.0002 \\
    59281.45623 & -18.8851 & 0.0048 & 0.0015 \\
    59303.45183 & -18.8889 & 0.0042 & 0.0073 \\
    59304.50004 & -18.8776 & 0.0043 & 0.0394 \\
    59305.56791 & -18.8411 & 0.0043 & 0.0138 \\
    59328.45107 & -18.8446 & 0.0042 & -0.0008 \\
    59329.43209 & -18.9104 & 0.0042 & -0.0121 \\
    59336.40591 & -18.8868 & 0.0056 & 0.0263 \\
    59337.46274 & -18.898 & 0.0046 & 0.0007 \\
    59340.39023 & -18.8957 & 0.0043 & 0.0225 \\
    59347.34985 & -18.8854 & 0.0044 & 0.0101 \\
    59348.45563 & -18.8869 & 0.0044 & 0.0134 \\
    59362.45989 & -18.9015 & 0.0041 & -0.0016 \\
    59392.37132 & -18.8889 & 0.0042 & 0.0195 \\
    59396.37382 & -18.8762 & 0.0049 & 0.0264 \\
    59532.6603 & -18.8697 & 0.0045 & 0.0153 \\
    59549.69237 & -18.914 & 0.0044 & 0.0317 \\
    59553.67898 & -18.8947 & 0.0048 & 0.0367 \\
    59561.64096 & -18.8762 & 0.0045 & -0.0102 \\
    59562.64694 & -18.8377 & 0.0043 & 0.008 \\
    59564.64881 & -18.896 & 0.0044 & 0.031 \\
    59565.6739 & -18.8801 & 0.0044 & 0.0167 \\
    59566.694 & -18.846 & 0.0045 & 0.0206 \\
    59568.64331 & -18.8907 & 0.0044 & 0.0109 \\
    59569.68166 & -18.8597 & 0.0044 & 0.0128 \\
    59570.62985 & -18.8421 & 0.0044 & 0.0068 \\
    59586.6029 & -18.894 & 0.0049 & 0.031 \\
    59590.6365 & -18.8931 & 0.0055 & 0.0276 \\
    59598.60044 & -18.8624 & 0.0049 & -0.0038 \\
    59605.52642 & -18.8963 & 0.0041 & 0.0025 \\
    59606.52812 & -18.8585 & 0.0044 & -0.0025 \\
    59609.63379 & -18.8675 & 0.0046 & -0.003 \\
    59610.60665 & -18.8436 & 0.0044 & 0.008 \\
    59621.49835 & -18.8536 & 0.0044 & -0.0095 \\
    59628.39137 & -18.8447 & 0.0052 & 0.0025 \\
    59660.4856 & -18.8876 & 0.0044 & -0.0382 \\
    59662.55404 & -18.852 & 0.0044 & 0.0103 \\
    59663.49547 & -18.9 & 0.0044 & 0.003 \\
    59684.46197 & -18.8148 & 0.0045 & 0.0341 \\
    59685.48672 & -18.8765 & 0.0045 & 0.0453 \\
    59686.54289 & -18.8943 & 0.0044 & 0.019 \\
    59732.37275 & -18.8732 & 0.0044 & 0.0075 \\
    59749.39015 & -18.8607 & 0.0055 & -0.0128 \\
    \hline                  
    \end{tabular}
\end{table}

\begin{table}[h!]
\small
\caption{SOPHIE RV and BIS time series for TOI-1273.}             
\label{tab:1271_rvs}      
\centering          
\begin{tabular}{l l l l }     
\hline\hline       
Time (RJD) & RV ($ms^{-1}$) & $\sigma_{\mathrm{RV}}$ ($ms^{-1}$) & BIS ($ms^{-1}$) \\
\hline                    
58887.60797 & -43.7134 & 0.0061 & 0.0194 \\
58916.47664 & -43.7684 & 0.0091 & 0.0177 \\
58917.69296 & -43.7675 & 0.007 & -0.021 \\
58918.52127 & -43.7312 & 0.0061 & -0.0006 \\
58919.564 & -43.7317 & 0.0097 & -0.0693 \\
59006.43351 & -43.7241 & 0.0049 & 0.0023 \\
59016.42816 & -43.7102 & 0.0065 & -0.039 \\
59017.45233 & -43.7216 & 0.006 & -0.0122 \\
59018.44308 & -43.7682 & 0.0057 & 0.0239 \\
59019.42139 & -43.7704 & 0.0051 & -0.0044 \\
59020.42867 & -43.7383 & 0.007 & 0.0246 \\
59056.37566 & -43.7588 & 0.0047 & -0.005 \\
59057.40103 & -43.7449 & 0.0039 & -0.016 \\
59058.35427 & -43.7246 & 0.0045 & -0.0002 \\
59059.36318 & -43.7301 & 0.0047 & -0.0218 \\
59060.3913 & -43.7657 & 0.004 & 0.0047 \\
59061.36421 & -43.7588 & 0.004 & -0.0018 \\
59062.36612 & -43.7381 & 0.0045 & 0.0168 \\
59081.34487 & -43.7223 & 0.0043 & 0.013 \\
59083.32672 & -43.775 & 0.004 & 0.0116 \\
59107.30036 & -43.7707 & 0.0064 & 0.0601 \\
59139.24583 & -43.782 & 0.0045 & -0.0117 \\
59140.2521 & -43.7452 & 0.0043 & -0.0021 \\
59141.24915 & -43.728 & 0.0047 & 0.0004 \\
59149.24854 & -43.7615 & 0.0091 & -0.0054 \\
59197.71414 & -43.7166 & 0.0039 & 0.0117 \\
59245.61019 & -43.7788 & 0.006 & 0.0123 \\
59247.72159 & -43.7247 & 0.0042 & -0.0004 \\
59248.6658 & -43.7171 & 0.0041 & 0.0167 \\
59270.57337 & -43.7374 & 0.0041 & 0.0232 \\
59273.66344 & -43.7624 & 0.0072 & -0.0315 \\
59275.55137 & -43.7283 & 0.0041 & -0.0091 \\
59278.55668 & -43.7879 & 0.0057 & 0.0227 \\
59279.59967 & -43.7529 & 0.0058 & 0.0082 \\
59280.62091 & -43.7203 & 0.0047 & 0.0286 \\
59303.46737 & -43.7274 & 0.0041 & 0.0067 \\
59304.51413 & -43.7263 & 0.004 & 0.0085 \\
59306.54667 & -43.774 & 0.0045 & -0.0135 \\
59327.50658 & -43.7152 & 0.0043 & 0.0117 \\
59329.47453 & -43.7751 & 0.0049 & 0.0172 \\
59404.40218 & -43.7568 & 0.0064 & 0.0054 \\
59561.70916 & -43.7554 & 0.0047 & 0.0115 \\
59563.71468 & -43.7232 & 0.0047 & 0.0275 \\
59565.71437 & -43.768 & 0.0045 & 0.0052 \\
59567.70234 & -43.7269 & 0.0041 & 0.0186 \\
59568.7174 & -43.723 & 0.0043 & 0.0087 \\
59569.6996 & -43.7594 & 0.0045 & 0.0076 \\
59570.69963 & -43.759 & 0.0049 & -0.0022 \\
59628.55175 & -43.7488 & 0.0068 & 0.023 \\
59630.6354 & -43.7975 & 0.0093 & 0.0165 \\
59660.57446 & -43.694 & 0.0052 & 0.0033 \\
59662.57218 & -43.7603 & 0.0044 & 0.0123 \\
59663.51279 & -43.7688 & 0.0043 & -0.0216 \\
59683.51875 & -43.7194 & 0.0046 & 0.0259 \\
59686.63687 & -43.7614 & 0.0087 & 0.0486 \\
59715.42352 & -43.7264 & 0.0043 & -0.0087 \\
59771.39786 & -43.741 & 0.0058 & -0.015 \\
59783.4065 & -43.7572 & 0.0058 & 0.0004 \\
59785.43504 & -43.7288 & 0.0047 & -0.0059 \\
59787.39647 & -43.7698 & 0.0071 & -0.0258 \\ 
\hline                  
\end{tabular}
\end{table}

\pagebreak

\onecolumn
\section{Prior distributions and posterior values}
\label{apx:priors}
In this appendix we present the prior distributions used in each model and the resulting values from the MCMC sampling of the posterior distributions for the fitted and derived parameters.
\begin{table}[h!]
\caption{Prior distributions and the resulting values from the posterior distributions for all the stellar and planetary parameters of the models.}     
\centering     
\begin{tabular}{l c c c c}    
\hline\hline       
 & \multicolumn{2}{c}{TOI-1199\:b} & \multicolumn{2}{c}{TOI-1273\:b} \\
\hline      
Parameter & Prior & Posterior & Prior & Posterior \\
\hline
\multicolumn{5}{l}{\textit{Stellar parameters}} \\
    $M_{\star}$ (\msol)  & $\mathcal{BN}(1.23, 0.07)$ & $1.228\pm0.067$ & $\mathcal{BN}(1.06, 0.06)$ & $1.030\pm0.058$ \\ 
    $R_{\star}$ (\RS)  & $\mathcal{BN}(1.45, 0.03)$ & $1.452\pm0.029$ &  $\mathcal{BN}(1.06, 0.02)$ & $1.069\pm0.019$ \\ 
    TESS q1 & $\mathcal{N}(0, 5)$ & $0.24\pm0.21$ & $\mathcal{N}(0, 5)$ & $0.49\pm0.32$ \\
    TESS q2 & $\mathcal{N}(0, 5)$ & $0.09\pm0.23$ & $\mathcal{N}(0, 5)$ & $0.12\pm0.37$ \\
    KeplerCam$^B$ q1 & $\mathcal{N}(0, 5)$ & $0.80\pm0.36$ & $\mathcal{N}(0, 5)$ & $0.72\pm0.37$ \\
    KeplerCam$^B$ q2 & $\mathcal{N}(0, 5)$ & $0.10\pm0.38$ & $\mathcal{N}(0, 5)$ & $0.03\pm0.42$ \\
    KeplerCam$^z$ q1 & $\mathcal{N}(0, 5)$ &  $1.33\pm0.28$ & -- & -- \\
    KeplerCam$^z$ q2 & $\mathcal{N}(0, 5)$ & $-0.52\pm0.28$ & -- & -- \\
    MuSCAT2$^g$ q1 & -- & -- & $\mathcal{N}(0, 5)$ & $0.45\pm0.32$ \\
    MuSCAT2$^g$ q2 & -- & -- & $\mathcal{N}(0, 5)$ & $0.14\pm0.36$ \\
    MuSCAT2$^i$ q1 & -- & -- & $\mathcal{N}(0, 5)$ & $0.99\pm0.41$ \\
    MuSCAT2$^i$ q2 & -- & -- & $\mathcal{N}(0, 5)$ & $-0.14\pm0.43$ \\
    MuSCAT2$^r$ q1 & -- & -- & $\mathcal{N}(0, 5)$ & $1.73\pm0.23$ \\
    MuSCAT2$^r$ q2 & -- & -- & $\mathcal{N}(0, 5)$ & $-0.78\pm0.22$ \\
    MuSCAT2$^{zs}$ q1 & -- & -- & $\mathcal{N}(0, 5)$ & $0.27\pm0.23$ \\
    MuSCAT2$^{zs}$ q2 & -- & -- & $\mathcal{N}(0, 5)$ & $0.05\pm0.24$ \\
\hline
\multicolumn{5}{l}{\textit{Planetary parameters}} \\
    $P$ (days) & $\log \mathcal{N}(1.30, 1)$ & $3.671463\pm0.000003$ & $\log \mathcal{N}(1.53, 1)$ & $4.631296\pm0.000003$ \\ 
    $T_0$ (TBJD) & $\mathcal{N}(2420.5,1)$ & $2420.5376\pm0.0004$ & $\mathcal{N}(1712.3, 1)$ & $1712.3468\pm0.0004$ \\ 
    $b$ & $\mathcal{U}(0,1+R_{p}/R_{\star})$ & $0.849\pm0.016$ & $\mathcal{U}(0,1+R_{p}/R_{\star})$ & $0.958\pm0.030$ \\ 
    $K$ (\ms) & $\log \mathcal{N}(3.34, 2)$ & $27.5\pm2.1$& $\log \mathcal{N}(3.27, 2)$ & $26.7\pm1.5$ \\ 
    $\sqrt{e} \sin{\omega}$ &  $\mathcal{U}(-1,1)$ & $-0.01\pm0.12$ & $\mathcal{U}(-1,1)$ & $-0.18\pm0.11$ \\ 
    $\sqrt{e} \cos{\omega}$ & $\mathcal{U}(-1,1)$ & $0.06\pm0.14$ & $\mathcal{U}(-1,1)$ & $0.04\pm0.14$ \\ 
    $R_{p}/R_{\star}$  & $\log \mathcal{N}(-2.79, 1)$ & $0.0663\pm0.0012$ & $\log \mathcal{N}(-2.66, 1)$ & $0.096\pm0.021$ \\ 
    $M$\,(\MJ)  & (derived) & $0.239\pm0.020$ & (derived) & $0.222\pm0.015$ \\ 
    $R$\,(\RJ)  & (derived) & $0.938\pm0.025$ & (derived) & $0.99\pm0.22$ \\ 
    $e$  & (derived) & $0.030\pm0.029$ & (derived) & $0.055\pm0.032$ \\ 
    $a$\,(AU)  & (derived) & $0.04988\pm0.00091$ & (derived) & $0.0549\pm0.0010$ \\ 
    $\rho~$ (\gcc)  & (derived) & $0.358\pm0.041$ & (derived) & $0.28\pm0.11$ \\ 
    $T_{\text{eq}}~$(K)  & (derived) & $1486\pm20$ & (derived) & $1211\pm15$ \\ 
    $\omega$ (rad)  & (derived) & $1.0\pm1.7$ & (derived) & $1.9\pm2.4$ \\ 
\hline
\label{tab:priors}
\end{tabular}
\tablefoot{$\mathcal{N}(\mu, \sigma)$ stands for normal distribution, in the case of stellar mass and radius $\mathcal{BN}$ means bounded normal and both distributions are bounded between 0 and 3. $\mathcal{U}(a,b)$ refers to a uniform distribution between $a$ and $b$ and $\log \mathcal{N}(\mu,\sigma)$ to a log-normal distribution.}
\end{table}

\begin{table}[h!]
\caption{Prior distributions and the resulting values from the posterior distributions for all the instrumental parameters.}     
\centering     
\begin{tabular}{l c c c c}    
\hline\hline       
 & \multicolumn{2}{c}{TOI-1199\:b} & \multicolumn{2}{c}{TOI-1273\:b} \\
\hline      
Parameter & Prior & Posterior & Prior & Posterior \\
\hline
\multicolumn{5}{l}{\textit{Instrumental parameters}} \\
    RV jitter (\ms) & $\log \mathcal{N}(1.25, 1)$ & $10.2\pm1.2$ & $\log \mathcal{N}(1.38, 1)$ & $7.5\pm1.1$ \\
    RV trend 2 & $\mathcal{N}(0, 0.01)$ & $0.00003\pm0.00002$ & $\mathcal{N}(0, 0.01)$ & $0.000010\pm0.000015$ \\
    RV trend 1 & $\mathcal{N}(0, 0.1)$ & $-0.006\pm0.006$ & $\mathcal{N}(0, 0.1)$ & $-0.0030\pm0.0045$ \\
    RV trend 0 & $\mathcal{N}(0, 1)$ & $0.63\pm0.93$ & $\mathcal{N}(0, 1)$ & $0.07\pm0.88$ \\
    TESS$^2$ offset & $\mathcal{N}(0, 5)$ & $0.136\pm0.040$ & $\mathcal{N}(0, 5)$ & $0.089\pm0.031$ \\
    TESS$^2$ jitter & $\mathcal{N}(0, 5)$ & $0.173\pm0.078$ & $\mathcal{N}(0, 5)$ & $0.110\pm0.049$ \\
    TESS$^{30}$ offset &  $\mathcal{N}(0, 5)$ & $0.089\pm0.049$ & $\mathcal{N}(0, 5)$ & $0.095\pm0.067$ \\
    TESS$^{30}$ jitter & $\mathcal{N}(0, 5)$ & $0.125\pm0.065$ & $\mathcal{N}(0, 5)$ & $0.097\pm0.057$ \\
    KeplerCam$^B$ offset & $\mathcal{N}(0, 5)$ & $0.12\pm0.14$ & $\mathcal{N}(0, 5)$ & $0.05\pm0.14$ \\
    KeplerCam$^B$ jitter & $\mathcal{N}(0, 5)$ & $1.18\pm0.12$ & $\mathcal{N}(0, 5)$ & $0.82\pm0.17$ \\
    KeplerCam$^z$ offset & $\mathcal{N}(0, 5)$ & $0.09\pm0.17$ & -- & -- \\
    KeplerCam$^z$ jitter & $\mathcal{N}(0, 5)$ & $1.55\pm0.13$ & -- & -- \\
    MuSCAT2$^g$ offset & -- & -- & $\mathcal{N}(0, 5)$ & $0.26\pm0.27$ \\
    MuSCAT2$^g$ jitter & -- & -- & $\mathcal{N}(0, 5)$ & $0.45\pm0.20$ \\
    MuSCAT2$^r$ offset & -- & -- & $\mathcal{N}(0, 5)$ & $-0.09\pm0.34$ \\
    MuSCAT2$^r$ jitter & -- & -- & $\mathcal{N}(0, 5)$ & $0.78\pm0.36$ \\
    MuSCAT2$^i$ offset & -- & -- & $\mathcal{N}(0, 5)$ & $-0.10\pm0.27$ \\
    MuSCAT2$^i$ jitter & -- & -- & $\mathcal{N}(0, 5)$ & $1.18\pm0.51$ \\
    MuSCAT2$^{zs}$ offset & -- & -- & $\mathcal{N}(0, 5)$ & $0.25\pm0.25$ \\
    MuSCAT2$^{zs}$ jitter & -- & -- & $\mathcal{N}(0, 5)$ & $0.46\pm0.21$ \\
\hline
\label{tab:priors2}
\end{tabular}
\tablefoot{$\mathcal{N}(\mu, \sigma)$ stands for normal distribution and $\log \mathcal{N}(\mu,\sigma)$ for a log-normal distribution. TESS$^2$ and TESS$^{30}$ represents the 2-min and 30-minute cadence data, respectively.}
\end{table}

\FloatBarrier
\section{Posterior distribution plots}
\label{apx:cornerplots}
In this this appendix we show the MCMC samples through corner plots for the posterior distributions of the planetary parameters.
\begin{figure*}[h!]
    \label{fig:corner1199}
    \centering
    \includegraphics[width=1.0\hsize]{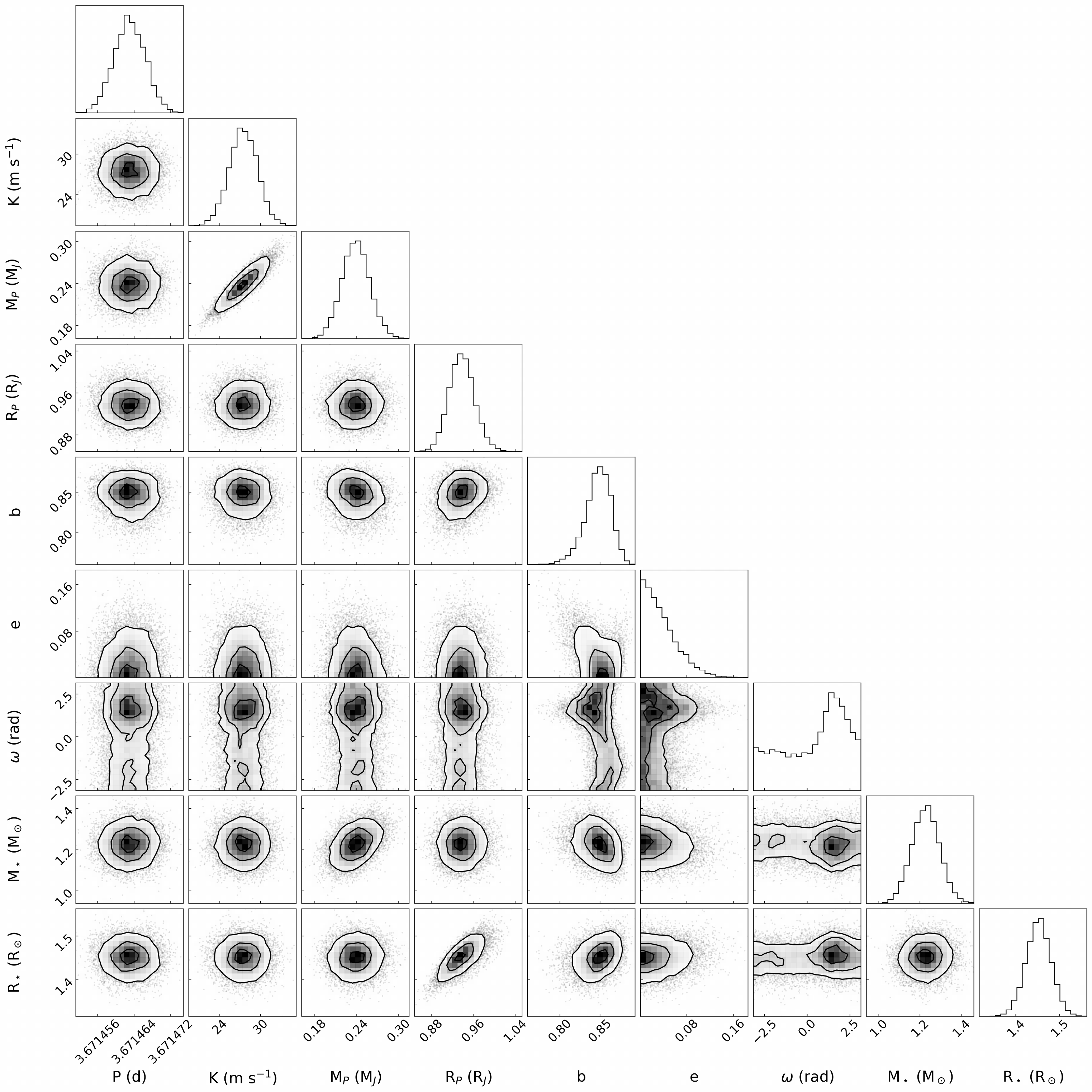}
      \caption{Results from the posterior distribution sampling of TOI-1199, the shape of the distributions, and the correlations between parameters. The contour levels show the 16th, 50th, and 84th quantiles.
      }
\end{figure*}
\begin{figure*}[h!]
    \label{fig:corner1273}
    \centering
    \includegraphics[width=1.0\hsize]{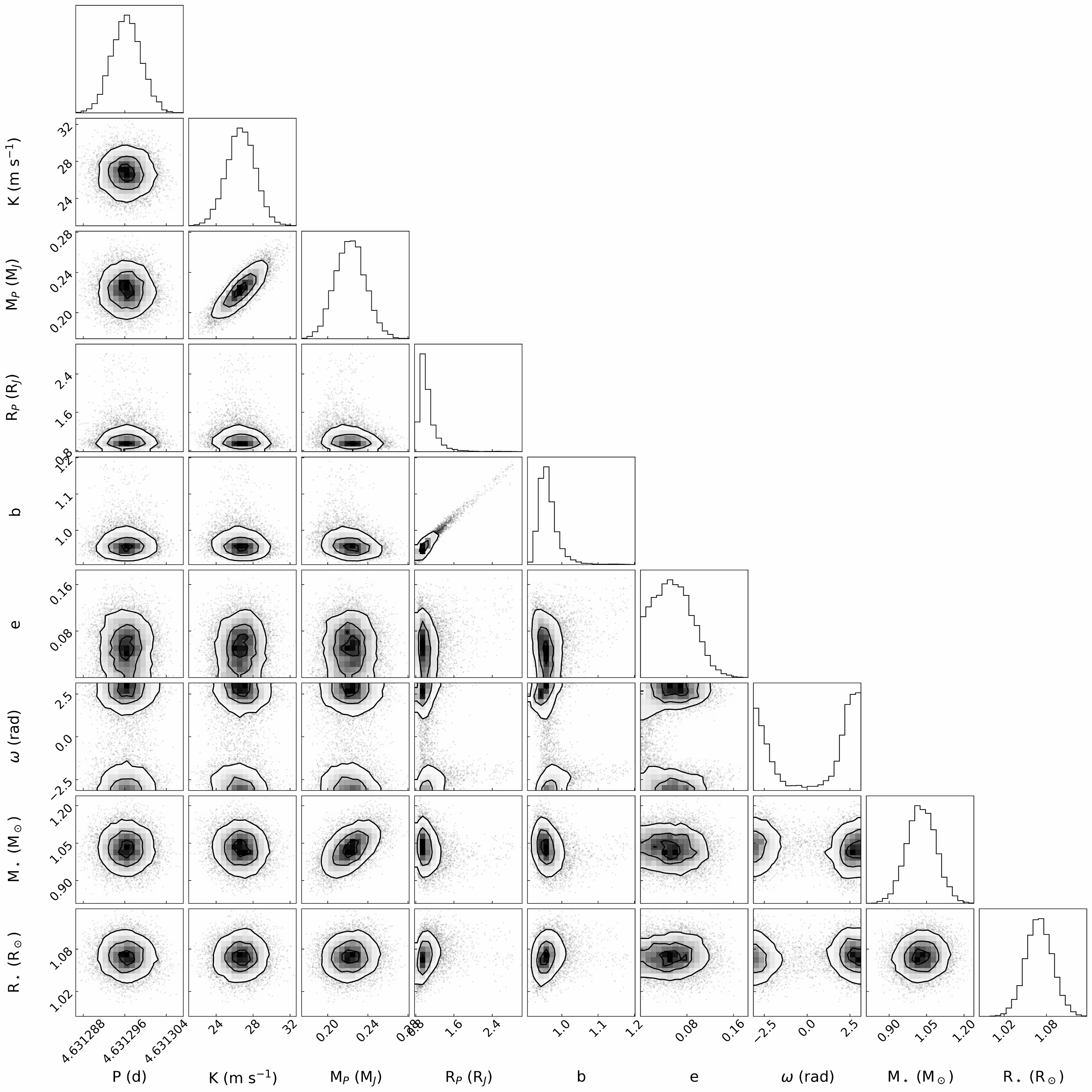}
      \caption{Results from the posterior distribution sampling of TOI-1273, the shape of the distributions; and the correlations between parameters The contour levels show the 16th, 50th, and 84th quantiles. A bimodal distribution appears for the argument of periastron; given the low eccentricity, this parameter is hard to constrain.
      }
\end{figure*}

\FloatBarrier
\section{Additional light curves}
\label{apx:extralcs}
In this appendix we show the ground light curves that were not used in the model.

\begin{figure*}[h!]
\centering
\includegraphics[width=16.4cm,clip]{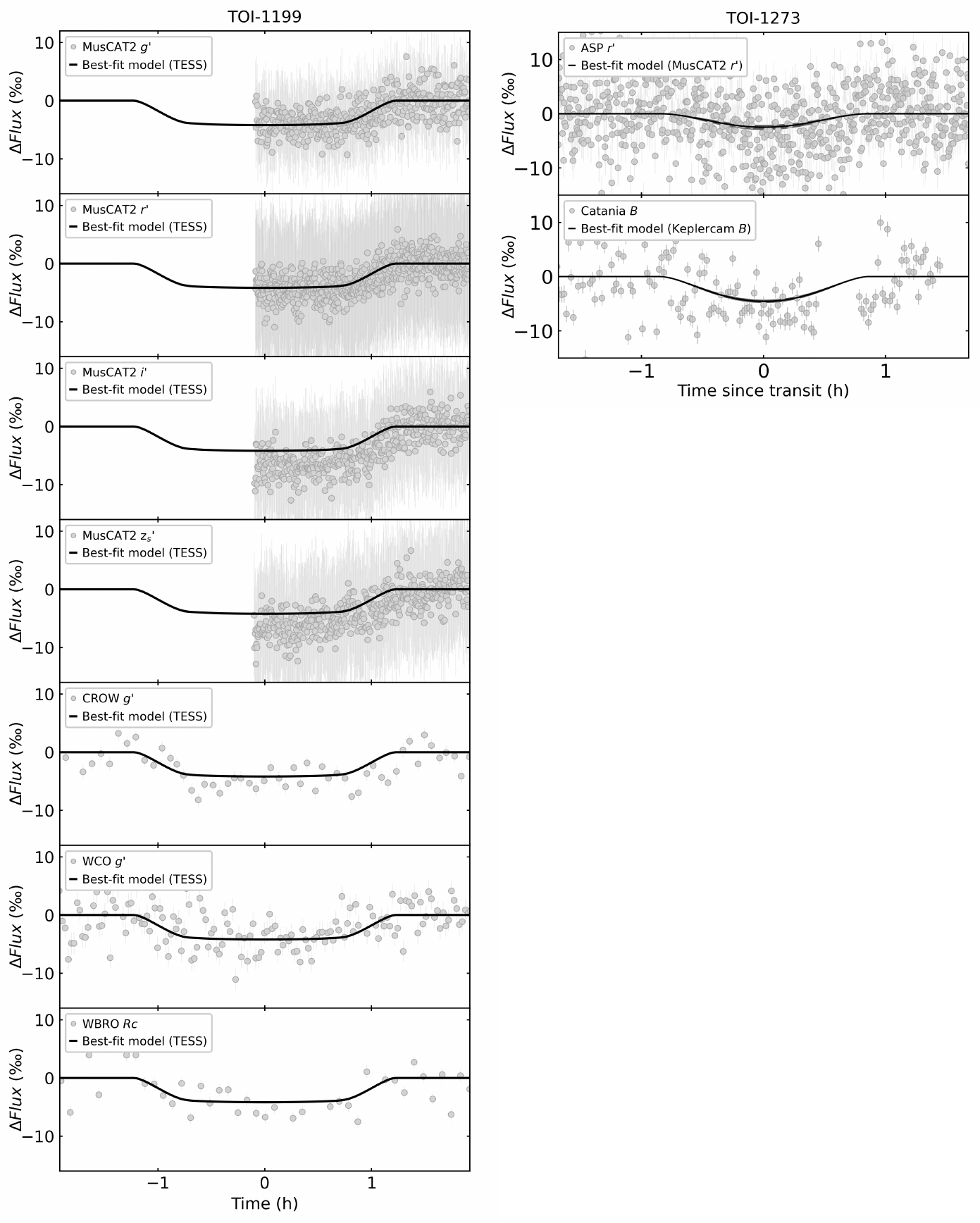}
\caption{Ground light curves not used in the modeling. \textit{Left panel:} MuSCAT2, CROW, WCO, and WBRO photometry for TOI-1199\:b. The best-fit model overplotted corresponds to the TESS 2-min light curve. \textit{Right panel:} ASP and Catania photometry for TOI-1273\:b. The best-fit model corresponds to different instruments but are in the same band.}
\label{fig:extralcs}
\end{figure*}
\end{appendix}
\end{document}